\documentclass[12pt,preprint]{aastex}
\usepackage{amssymb}

\shorttitle{IR Properties of LSB Galaxies}
\shortauthors{Rahman, Howell, Buckalew, Helou, \& Mazzarella}
 
\begin{document}
 
\title{Exploring Infrared Properties of Giant Low Surface Brightness Galaxies}
 
\author{Nurur Rahman\altaffilmark{1}, Justin H. Howell, George Helou, 
Joseph M. Mazzarella, and Brent Buckalew}
\affil{Infrared Processing and Analysis Center (IPAC)/Caltech, \\
Mail Code 100-22, 770 S. Wilson Avenue, \\
Pasadena, CA 91125, USA}
\email{nurur@ipac.caltech.edu}
 
\altaffiltext{1}{National Research Council (NRC) Postdoc Fellow}
\begin{abstract}
We present analysis of {\it Spitzer} Space Telescope observations of the three 
low surface brightness (LSB) optical giant galaxies Malin 1, UGC 6614 and UGC 
9024. Mid- and far-infrared morphology, spectral energy distributions, and 
integrated colors are used to derive the dust mass, dust-to-gas mass ratio, 
total infrared luminosity, and star formation rate (SFR). We also investigate 
UGC 6879, which is intermediate between high surface brightness (HSB) and LSB 
galaxies. 
The 8 $\mu$m images indicate that polycyclic aromatic hydrocarbon (PAH) 
molecules are present in the central regions of all three metal-poor LSB 
galaxies. The diffuse optical disks of Malin 1 and
UGC 9024 remain undetected at mid- and far-infrared wavelengths. The
dustiest of the three LSB galaxies, UGC 6614, has infrared morphology that
varies significantly with wavelength; 160 $\mu$m  (cool) dust
emission is concentrated in two clumps on the NE and NW sides of a distinct
ring seen in the 24 and 8 $\mu$m images (and a broken ring at 70 $\mu$m)
at a radius of $\sim$40\arcsec \ (18 kpc) from the galaxy center. The 8 and 
24 $\mu$m \ emission is co-spatial with H$\alpha$ emission previously 
observed in the outer ring of UGC 6614. The estimated dust-to-gas ratios, 
from less than $10^{-3}$ to $10^{-2}$, support previous indications that the 
LSB galaxies are relatively dust poor compared to the HSB galaxies. 
The total infrared luminosities are approximately 1/3 to 1/2 the blue band 
luminosities, suggesting that old stellar populations are the primary source 
of dust heating in these LSB objects. The SFR estimated from the infrared data 
ranges $\sim$$\rm 0.01-0.88~M_\odot~yr^{-1}$, consistent with results from 
optical studies.
\end{abstract}
\keywords{galaxies: spirals - galaxies: dust - galaxies: ISM - galaxies: 
structure - galaxies: individual (Malin 1, UGC 6614, UGC 6879, UGC 9024)}
\section{Introduction}
The unprecedented imaging and spectroscopic sensitivity and higher spatial 
resolution of the {\it Spitzer} Space Telescope ({\it SST}; Werner et al. 2004) 
compared to past IR missions such as the Infrared  Astronomical Satellite 
({\it IRAS}; Neugebauer et al. 1984) and  the Infrared  Space Observatory 
({\it ISO}; Kessler et al. 1996) provide a unique opportunity to probe the 
inter-stellar medium (ISM) of optically faint sources such as low surface 
brightness (LSB) galaxies. The goal of this study is to use the {\it Spitzer} 
data to analyze the IR properties of three LSB optical giants: Malin 1, UGC 
6614, and UGC 9024. This is the first time we are able to view these galaxies 
in $\sim$3-160 $\mu$m wavelength range. 

The LSB galaxies are usually defined as diffuse spiral disks with low $B$-band 
central surface brightness (e.g. $\mu_{B,0} \geq 23$ mag arcsec$^{-2}$; 
Bothun et al. 1997; Impey \& Bothun 1997). These galaxies are either blue 
(B-V $\lesssim 0.5$) or red (B-V $\gtrsim 0.8$) in color (O'Neil et al. 1997), 
metal poor ([O/H] $\lesssim 1/3 Z_{\odot}$; McGaugh 1994; de Blok \& van der 
Hulst 1998b; de Naray et al. 2004), rich in neutral hydrogen (H I) (Schombert 
et al. 1992; O'Neil et al. 2004), deficient in H II emission (McGaugh et al. 
1995; de Naray et al. 2004), and have low star formation rate (SFR) 
$\lesssim 0.1$ M$_{\odot}$ yr$^{-1}$ (van den Hoek et al. 2000; Burkholder et 
al. 2001). 
The majority of these galaxies lack molecular (CO) gas (Schombert et al. 1990; 
de Blok \& van der Hulst 1998a); only a handful have been reported to have 
molecular emission (O'Neil \& Schinnerer 2004; Matthews et al. 2005; Das et al. 
2006).The observed properties suggest that LSB disks are relatively unevolved 
systems and may have a different evolutionary history compared to their high 
surface brightness (HSB) counterparts (McGaugh 1994; van den Hoek et al. 2000; 
Vallenari et al. 2005; Zackrisson et al. 2005). 
 
Most of our knowledge regarding the composition and structure of the ISM of 
LSB spirals comes from optical (Impey et al. 1996) and H I surveys (O'Neil et 
al. 2004). These surveys have demonstrated that the neutral hydrogen is by far 
the dominant component of the ISM in these galaxies ($\sim$95\% by mass; 
Matthews 2005). While we have improved understanding of the gaseous component 
of the ISM from decade long H I surveys, our knowledge of the inter-stellar 
dust, the component of the ISM radiating from mid-IR ($\sim$8 $\mu$m) through 
sub-millimeter ($\sim$850 $\mu$m) wavelengths, is still very limited. Because 
of the scarcity of information in this wavelength range, complementary 
observational facts such as low metal abundance (McGaugh 1994), strong 
similarities in optical and near-IR morphology (Bergvall et al. 1999; Bell et 
al. 2000), transparency of the stellar disks (O'Neil et al. 1998; Matthews et 
al. 1999), and deficiency in molecular emission (Schombert et al. 1990; de Blok 
\& van der Hulst 1998a) have been used to probe the ISM of LSB galaxies. All 
these observations lead to a general consensus that the LSB disks are deficient 
in dust and molecular gas.  
 
Given that LSB spirals comprise $\sim$$50\%$ of the local disk galaxy population 
(McGaugh et al. 1995b), they deserve equal attention as their HSB cousins. 
To develop a consistent picture of the local galaxy populations it is therefore 
necessary to probe each population at all wavelength regimes as have been done 
in most cases for HSB galaxies. Previous long wavelength studies on LSB galaxies 
involved a few cases in sub-millimeter and millimeter wavelengths (de Blok \& 
van der Hulst 1998b; Pickering \& van der Hulst 1999; O'Neil \& Schinnerer 2004; 
Matthews et al. 2005; Das et al. 2006). Hoeppe et al. (1994) made the first 
attempt to investigate long wavelength (60 $\mu$m, 100 $\mu$m, and 20 cm) 
properties of LSB dwarfs; however, no study has been made in the mid-IR and 
far-IR for LSB disks. In this study we have made the first attempt to explore 
$\sim$3-160 $\mu$m properties of three LSB giants: Malin 1, UGC 6614, and UGC 
9024, recently observed by the {\it SST}. We focus on the IR morphology to probe 
extent of dust in the ISM, on the SEDs for the IR energy budget and dust content, 
and on the IR colors to establish the dust temperature of the ISM. 
The organization of this paper is as follows: we describe observation and data 
reduction in section $\S2$ and present our results in section $\S3$. Discussions 
and conclusions are given in section $\S4$.
\section{Observation and Data Reduction}
We describe the Infrared Array Camera (IRAC; Fazio et al. 2004) and Multiband 
Imaging Photometer for {\it Spitzer} (MIPS; Rieke et al. 2004) imaging data for 
Malin 1, UGC 6614, and UGC 9024. These extended disk galaxies with 
radial scale length $\rm h_{r,R}>$ 5 kpc are observed as part of a larger 
guaranteed time observing program ({\it Spitzer} PID \#62). The program also 
includes two LSB galaxies (UGC 5675 and UGC 6151) with 2 kpc $<\rm h_{r,R}<$ 3 
kpc, an edge-on disk (UGC 6879) with $\rm h_{r,R} \sim$2.5 kpc, and a HSB 
dwarf (UGC 10445) with $\rm h_{r,R} \sim$1 kpc. The central brightness of UGC 
6879 is $\mu_{B,0} \sim 20.4$  mag arcsec$^{-2}$. A simple correction for 
inclination, using $\mu_{0, face-on} = \mu_{0, observed} + 2.5 \log (a/b)$ 
brings its central brightness to $\sim 21.71$ mag arcsec$^{-2}$. It is lower 
than the conventional choice ($\mu_0 \approx 23$ mag arcsec$^{-2}$) but close 
to the Freeman value ($\mu_0 \sim 21.65$ mag arcsec$^{-2}$; Freeman 1970). 
The central brightness of this galaxy falls in between the range of LSB and 
HSB galaxies and hence, we include it as a representative of the intermediate 
class. The properties of UGC 5675 and UGC 6151 will be explored in a 
forthcoming paper. The readers are refereed to Hinz et al. (2006) for an 
analysis of UGC 10445.

The IRAC 3.6, 4.5, 5.8, and 8 $\mu$m images and the MIPS 24, 70, and 160 
$\mu$m images were acquired, respectively, in the mapping and photometry 
modes. The IRAC images were reduced with the standard {\it Spitzer} Science 
Center data pipeline, and aligned, re-sampled, and combined into a mosaic 
image using the {\it Mopex}\footnote{http://ssc.spitzer.caltech.edu/postbcd/} 
software. The MIPS 24 $\mu$m data required the use of the self-calibration 
procedure described in the MIPS Data 
Handbook\footnote{http://ssc.spitzer.caltech.edu/mips/dh/} to remove latent 
image artifacts.  The corrected images were then combined into a mosaic 
using {\it Mopex}. Time filtering and column filtering was applied to the 
70 $\mu$m images using IDL routines created by D. Fadda. The filtered images 
were then combined using {\it Mopex}.  The 160 $\mu$m images were combined 
into a mosaic using {\it Mopex}. The IRAC spatial resolution is $\sim$2$^{''}$ 
for all bands. The MIPS spatial resolutions are 6$^{''}$, 18$^{''}$, and 
40$^{''}$ for the respective bands. 

Sky subtraction was carried out through the use of multiple sky apertures 
placed near the source which do not overlap with the faintest isophotes 
visible from the galaxy. For each galaxy we measured flux densities from 
the sky subtracted images within the aperture covering the entire galaxy.
The flux density contributed by foreground stars within a galaxy aperture 
was removed by measuring each such star in a small aperture and subtracting 
the result from the total flux within the galaxy aperture.
The calibration uncertainty in the IRAC flux densities is at the level 
of $\sim$10\% (Reach et al. 2005; Dale et al. 2005). Aperture corrections 
have been applied to all IRAC flux densities (T. H. Jarrett 2006; private 
communication). The MIPS flux density calibration uncertainties are 10\% at 
24 $\mu$m and 20\% at 70 and 160 $\mu$m. 

Near-IR (1.3, 1.7, and 2.2 $\mu$m) flux densities from the Two Micron All 
Sky Survey (2MASS; Jarrett et al. 2000), upper limits on {\it IRAS} flux 
densities, and those derived from the IRAC and MIPS bands are given in 
Table \ref{flux_table}. Basic properties of the galaxies obtained from the 
literature, the NASA/IPAC Extragalactic Database (NED), the Lyon-Meudon 
Extragalactic Database (LEDA), and derived in this study are summarized in 
Table \ref{basic_table}. The {\it IRAS} flux densities 
for UGC 6614 and UGC 6879 were computed using the SCANPI tool available from 
IRSA as linked via NED where {\it IRAS} flux density limits represent SCANPI's 
inband total measurement, $f_{\nu}(t)$. No {\it IRAS} detections were available 
for Malin 1 and UGC 9024. Distances for the galaxies are estimated from 
heliocentric radial velocity after correcting for the local group, infall into 
the Virgo cluster, the Great Attractor, and the Shapley supercluster following 
the Mould et al. (2000) flow model. 
\subsection{Contamination from Galactic Cirrus}
A basic concern about faint extragalactic sources with highly diffuse disk 
structure is confusion by foreground Galactic ``cirrus'' emission. This is 
especially critical in the case of far-IR cool sources (defined below) such 
as LSB galaxies. The far-IR ratio of typical local cirrus is 
$S_{60 \mu m}/S_{100 \mu m} \leq 0.2$ (Gautier 1986). UGC 6614 and UGC 6879 
are well above this limit (Fig. \ref{colo_colo}c). From its Galactic latitude 
$b \sim +22^{o}$ and the observed $S_{70 \mu m}/S_{160 \mu m}$ ratio, it is 
reasonably safe to assume that UGC 9024 has not been affected by cirrus 
emission. In the absence of far-IR information it is uncertain for  Malin 1. 
However, its high galactic latitude $b \sim +14^{\circ}$ can be used to argue 
against any cirrus contamination. 

We should also stress the fact that whether there is cirrus in the foreground 
is less significant as compared to whether the cirrus in the foreground varies 
on scales of the IRAC and MIPS field of view.  If its spatial variation is 
negligible across the IRAC and MIPS mosaic, it gets subtracted out as ``sky'' 
in the data reduction process.
\section{Results}
In this section we present IR morphology, SEDs, and IR colors of LSB galaxies. 
Using this information we estimate dust mass ($\rm M_d$), dust-to-(atomic) 
gas ratio ($\mathcal D$), total infrared luminosity ($\rm L_{TIR}$), and star 
formation rate (SFR). To obtain a qualitative assessment of the observed 
properties of LSB disks compared to their HSB counterparts, we take the 
{\it Spitzer} Infrared Nearby Galaxy Survey (SINGS; Kennicutt et al. 2003) 
sample as a representative sample of local HSB galaxies. This sample contains 
75 galaxies of various Hubble types as well as dwarfs and irregular galaxies, 
and thus making it a suitable reference for comparative analysis with LSB 
spirals. 
\subsection{Infrared Morphology}
The IR emission beyond $\sim$25 $\mu$m is dominated by the inter-stellar dust 
under various heating conditions. On the other hand, mid-IR ($\sim$5-25 $\mu$m) 
emission marks the transition from stellar photospheres to inter-stellar dust 
dominating the emission. Whereas the morphology of a galaxy at 3.6 and 4.5 
$\mu$m represents the stellar disk, at $\sim$5 $\mu$m and beyond it shows the 
structure of the ISM. The IRAC 3.6 and 4.5 $\mu$m bands are sensitive to the 
underlying stellar populations typically consisting of red giants and old 
stars. In some galaxies 3.6 $\mu$m band is also known to contain emission 
from a hot dust component (Bernard et al. 1994; Hunt et al. 2002; Lu et al. 
2003; Helou et al. 2004).
The hot dust is also visible in the other IRAC bands. While 3.6 $\mu$m is 
only sensitive to hot dust near the sublimation temperatures ($\sim$1000 K), 
the longer wavelength IRAC bands can detect dust at lower temperatures down 
to several hundred Kelvin.

The IRAC 5.8 and 8 $\mu$m bands are primarily sensitive to the PAH emission 
at 6.2, 7.7, and 8.6 $\mu$m  (Puget \& Leger 1989; Helou et al. 2000; Lu et 
al. 2003). The PAH is the hot component of the inter-stellar dust with 
effective temperature $\rm T_d > 100$ K stochastically excited to high energy 
levels by stellar photons. Stellar photospheric emission also contributes to 
these two IRAC wavebands. The fraction of stellar emission at 5.8 and 8 $\mu$m, 
respectively, are $\sim$40\% and $\sim$20\% (Helou et al. 2004; Pahre et al. 
2004). 

The emission detected by the MIPS bands are coming from dust grains with 
different size distributions. Very small grains ($\sim$1-10 nm) emit in the 
mid-IR region ($\gtrsim$15 $\mu$m), intermediate between thermal equilibrium 
and single-photon heating. Large, classical grains ($\sim$100-200 nm) are in 
thermal equilibrium with the radiation field and responsible for far-IR 
emission (Desert et al. 1990). To begin with we should bear in mind that the 
demarcation lines among various heating environments are {\it ad-hoc} and we 
assume the following effective color temperature ranges for large grains in 
thermal equilibrium: warm (40 K $\lesssim \rm T_d \lesssim$ 100 K), cool (20 
K $\lesssim \rm T_d \lesssim $ 30 K), cold (10 K $\lesssim \rm T_d \lesssim $ 
20 K), and very cold (10 K $\lesssim \rm T_d$). 

\begin{figure}
\epsscale{0.80}
\plotone{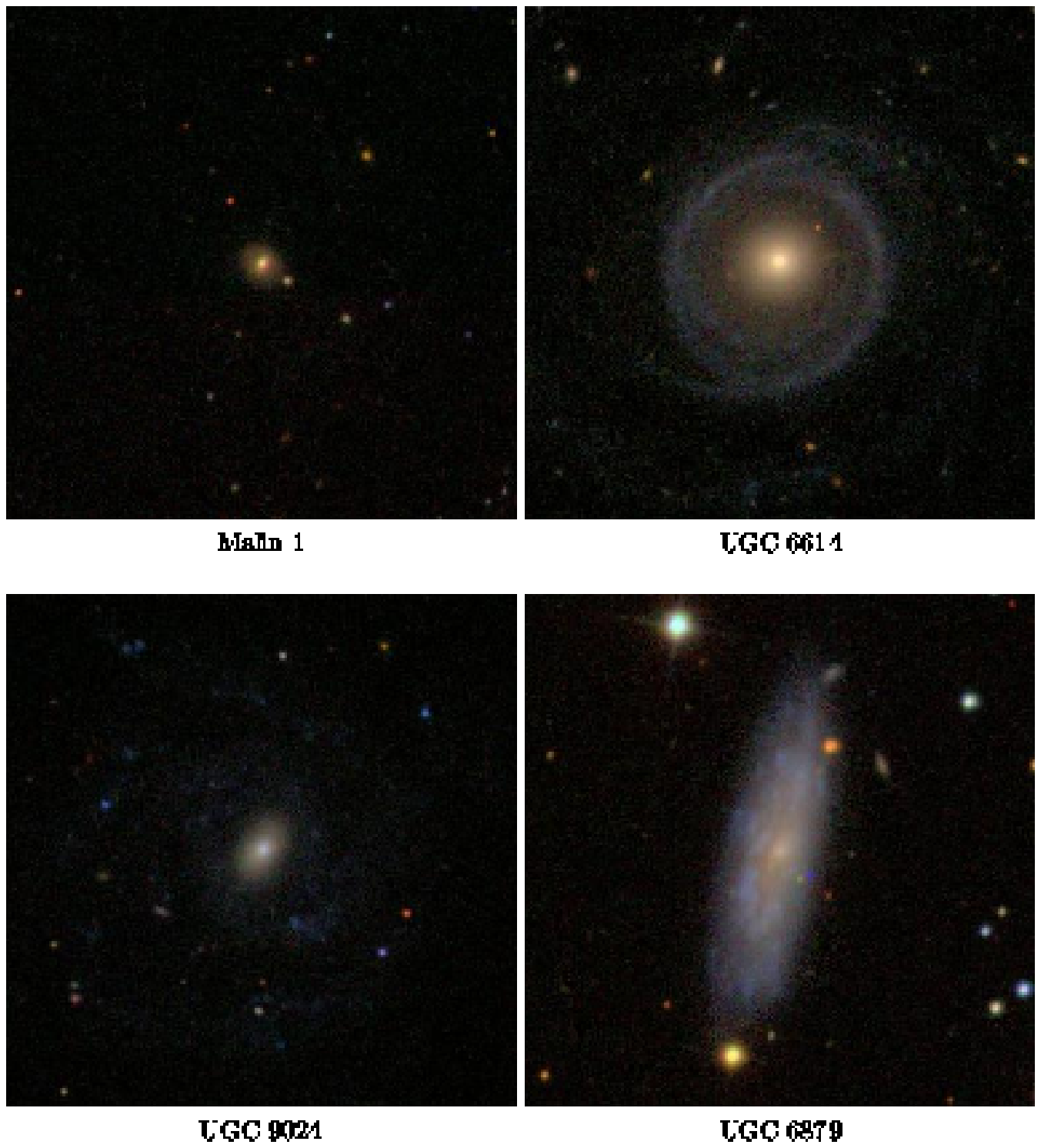}
\caption{\small
The Solan Digitial Sky Survey (SDSS) composite images of target galaxies. In each 
image north is up and east is to the left. The field of view is $2.5\arcmin \times 
2.5\arcmin$ with 0.4\arcsec \ resolution. Note that the diffuse disks of Malin 1 
and UGC 9024 are barely visible in these images. To help visualize these extended 
disks readers are referred to Barth (2007) for {\it I}-band {\it Hubble} image of 
Malin 1 and the following website for deeper {\it B}-band images of UGC 6614 and 
UGC 9024 (http://zebu.uoregon.edu/sb2.html).\label{optical}}
\end{figure}

\begin{figure}
\epsscale{0.80}
\plotone{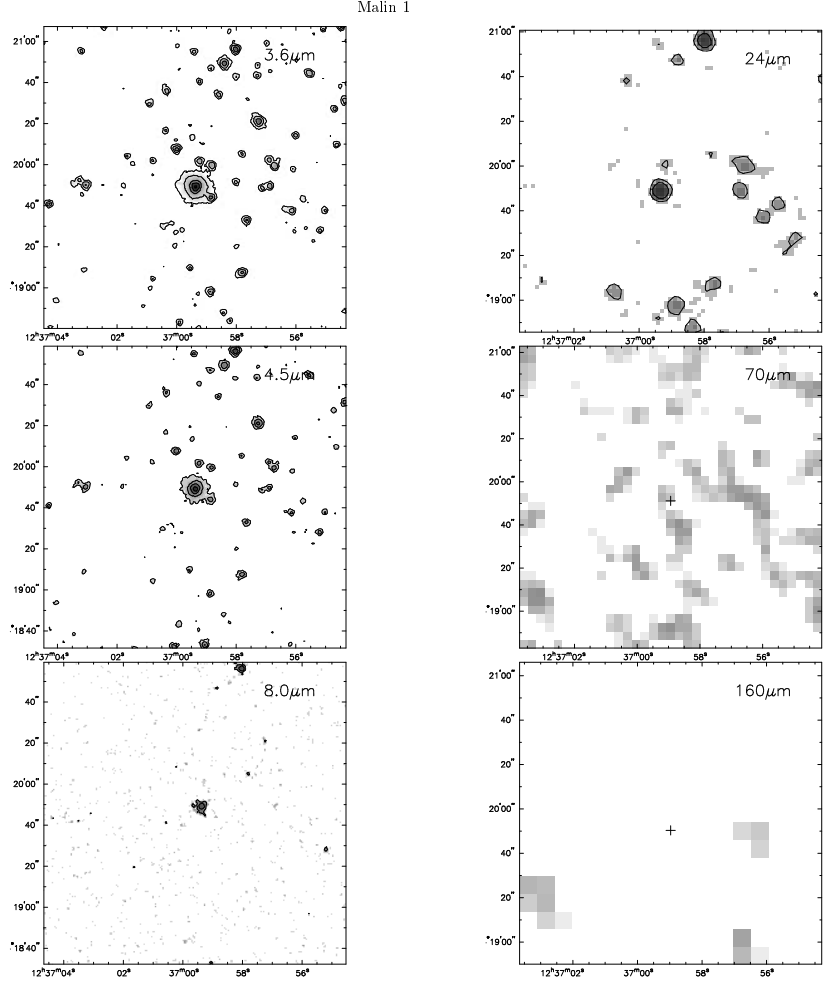}
\figcaption{\small
The {\it Spitzer} view of Malin 1. The IRAC 3.6, 4.5, and 8 $\mu$m images are at 
left and the MIPS 24, 70, and 160 $\mu$m images are at right. The IRAC 4.5 and 
8 $\mu$m images are shown without subtracting stellar photospheric emission. 
In each image north is up and east is to the left. The field of view is 
$2.5\arcmin \times 2.5\arcmin$ in all bands. Pixel sizes are $0.61\arcsec$ for 
the IRAC bands and $1.8\arcsec$, $4.0\arcsec$, and $8.0\arcsec$ for the MIPS 24, 
70, and 160 $\mu$m bands, respectively. 
Galaxies from Figs. \ref{ugc6614}-\ref{ugc6879} are presented in a similar 
manner. There is no detection of Malin 1 at 70 and 160 $\mu$m and hence the 
position of 24 $\mu$m peak emission is shown by the ``+'' sign at these bands. 
The contours represent surface brightness (MJy/Sr) with intervals of $\sqrt{10}$ 
where the lowest level is 4$\sigma$ above the background. See text for values of 
the lowest contour levels at different bands. \label{malin1}}
\end{figure}

\begin{figure}
\epsscale{0.80}
\plotone{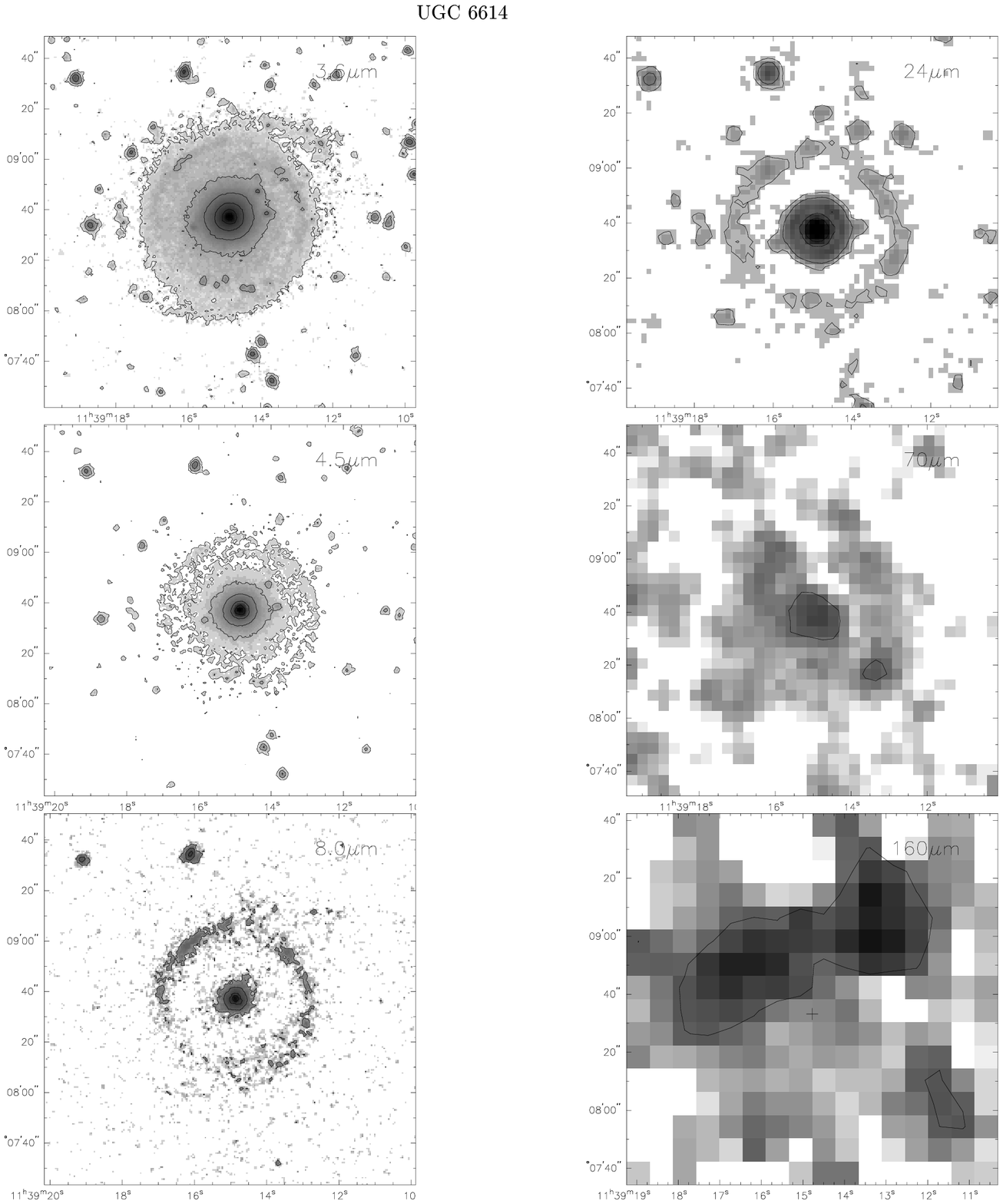}
\caption{\small
The {\it Spitzer} view of UGC 6614. The position of 24 $\mu$m peak emission is 
shown by the ``+'' sign at 160 $\mu$m image. \label{ugc6614}}
\end{figure}

\begin{figure}
\epsscale{0.80}
\plotone{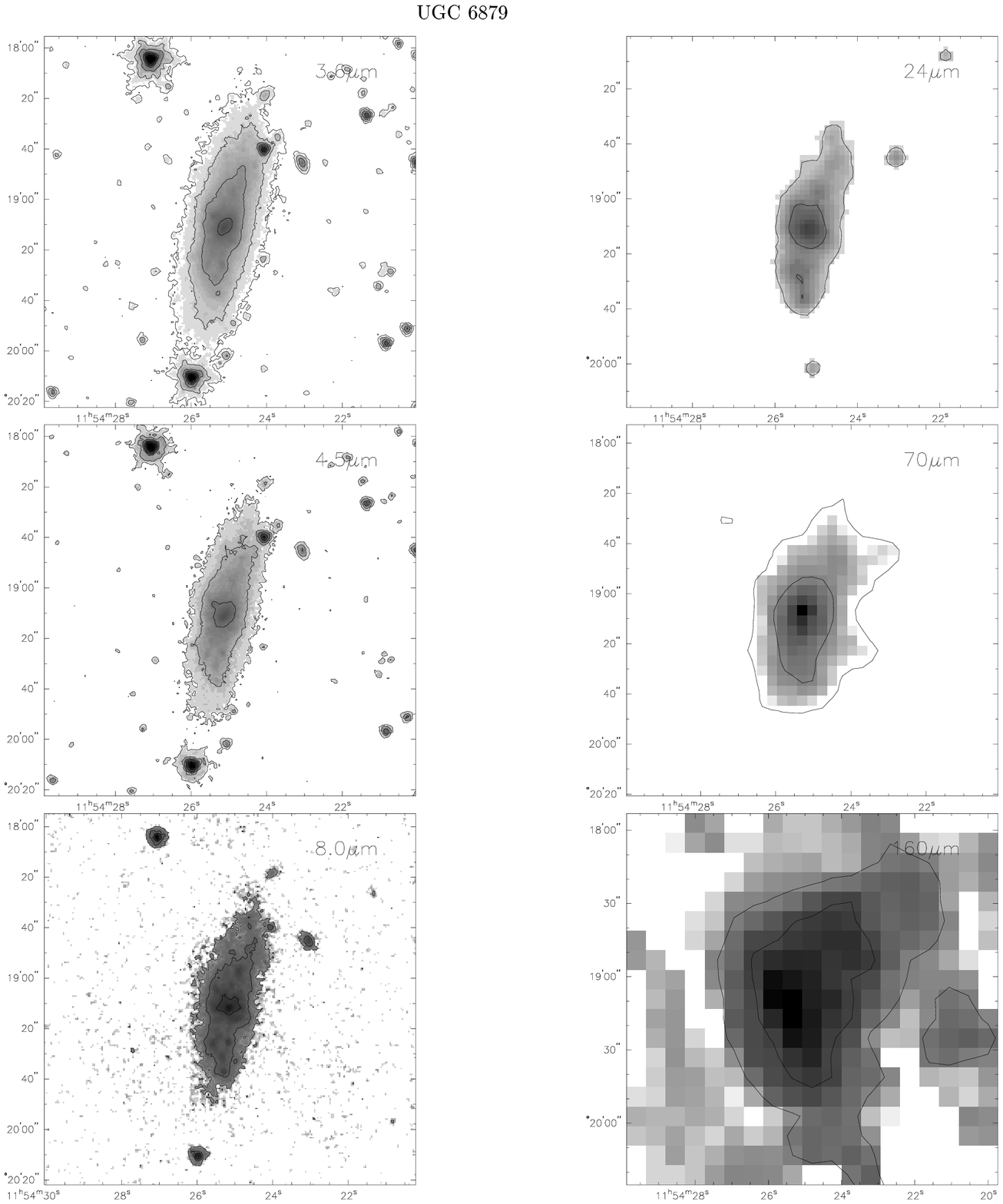}
\caption{\small
The {\it Spitzer} view of UGC 9024. The position of 24 $\mu$m peak emission is 
shown by the ``+'' sign at 160 $\mu$m image. \label{ugc9024}}
\end{figure}

\begin{figure}
\epsscale{0.80}
\plotone{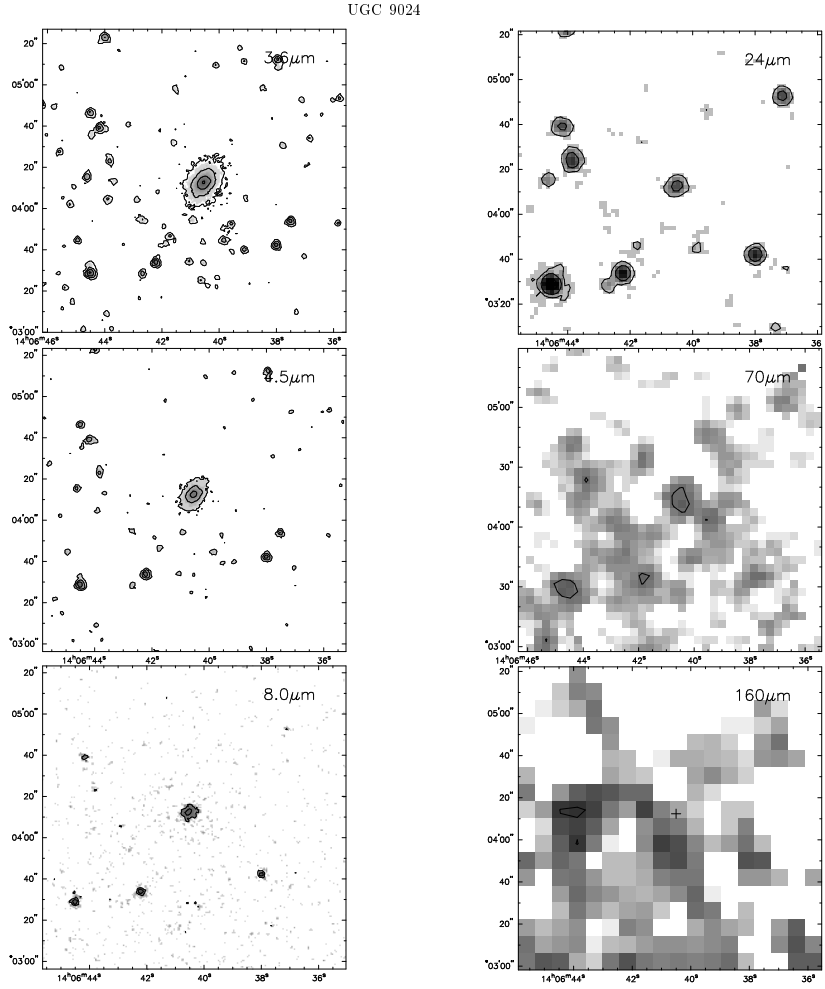}
\caption{\small
The {\it Spitzer} view of UGC 6879. The $B$-band central surface brightness 
$\mu_{B,0}$ of this galaxy is intermediate between LSB and HSB galaxies. 
\label{ugc6879}}
\end{figure}

The Solan Digital Sky Survey (SDSS) images of target galaxies are shown in 
Fig. \ref{optical}. The {\it Spitzer} images of the LSB galaxies 
are shown in Fig. \ref{malin1} (Malin 1), Fig. \ref{ugc6614} (UGC 6614), and 
Fig. \ref{ugc9024} (UGC 9024). UGC 6879 is shown in Fig. \ref{ugc6879}. Galaxy 
images are shown using Gaussian equalization (Ishida 2004). 
The images are oriented such that north is up and east is to the left. Malin 1 
is too faint to be detected by the MIPS 70 and 160 $\mu$m channels. The other 
three galaxies were detected by all IRAC and MIPS bands. We do not show 5.8 
$\mu$m images in these figures since the morphological appearances of each 
of these galaxies at 5.8 $\mu$m closely follow that at 8 $\mu$m. For all 
galaxies, the IRAC 4.5 and 8 $\mu$m images are shown without subtracting 
stellar photospheric emission. The contours represent surface brightness with 
intervals of $\sqrt{10}$ where the lowest level is 4$\sigma$ above the 
background. The lowest level of contours at different bands are 0.04 (3.6 
$\mu$m), 0.05 (4.5 $\mu$m), 0.30 (8 $\mu$m), 0.08 (24 $\mu$m), 1.6 (70 $\mu$m), 
and  2.1 (160 $\mu$m) expressed in MJy/Sr. 
 
The giant optical disks in Malin 1 and UGC 9024 appear as point sources in the 
IRAC images. While we detect the stellar bulges of these galaxies, their optically 
diffuse disks remain undetected long-ward of the IRAC bands. This suggests that the 
low surface brightness structures at larger radii might be photometrically distinct 
components rather than smoothed extensions of the normal inner disks (see Barth 
2007 for a discussion on Malin 1 based on {\it Hubble} data). That the disks 
appear in the $B$-band but are undetected at 3.6 $\mu$m  suggest that these disks 
have a small population of young stars rather than a large population of old stars. 
The bulge spectrum of Malin 1 is consistent with a predominantly old stellar 
population (Impey \& Bothun 1989). For both of these galaxies, the mid-IR emissions 
at 8 and 24 $\mu$m  are concentrated in the central few kpc, within a region of  
12\arcsec \ (20 kpc) radius for Malin 1 and 24\arcsec \ (5 kpc) radius 
for UGC 9024.

Undetected far-IR emission from the disk of Malin 1 implies that it contains 
either very cold ($\rm T_d <10$ K) cirrus-like dust emitting in the submillimeter 
and millimeter wavelengths or it lacks cold dust altogether and contains only 
neutral gas. For UGC 9024, 70 $\mu$m  emission comes from the central region 
but 160 $\mu$m emission is very hard to measure because of large scale diffuse 
emission in the field. This result in an upper limit of $S_{160 \ \mu m}<268$ 
mJy.

The optical morphology of UGC 6614 shows a massive bulge and spiral structure. 
A thin ring $\sim$40\arcsec \ from the core is prominent in H$\alpha$ (McGaugh 
et al. 1995). The 3.6 and 4.5 $\mu$m  emission is spread over the entire disk of 
this galaxy. The 3.6 $\mu$m image shows a discernible spiral arm pattern closely 
resembling the optical morphology. At 4.5 $\mu$m this feature disappears and 
the disk shrinks in radius showing only its inner region. This galaxy appears 
markedly different at 5.8 and 8 $\mu$m compared to the other IRAC bands. The 8 
$\mu$m morphology suggests that the PAH emission is coming from two distinct 
regions: the central bulge and an outer ring surrounding the bulge. The 8 $\mu$m 
morphology closely traces the H$\alpha$ image.
The MIPS 24 $\mu$m morphology is similar to the 8 $\mu$m PAH emission although 
the outer ring appears a bit more disjointed at this band. The dust emission at 
24 $\mu$m is coming mostly ($\sim$70\%) from the central disk.
The lower resolution image at 70 $\mu$m  indicates that only $\sim$25\% of the 
dust emission is coming from the central part of the galaxy, with the remaining 
$\sim$75\% emission co-spatial with the ring of radius $\sim$40\arcsec. 
Surprisingly a dumbbell shaped region is the dominant source ($\sim$90\%) of 160 
$\mu$m emission; these two peaks are located on the NE and NW sides of the ring.
The far-IR images of this galaxy show a small, localized region within the ring, 
SW from the center. Whether or not this region coincide with the spatial location 
showing the CO emission in this galaxy disk (Das et al. 2006) is not entirely 
clear. We will investigate this in the forthcoming paper. 

UGC 6879 is an edge on spiral with a red central part and a blue (optical) disk. 
This radial color gradient is perhaps related to greater concentrations of dust 
in the nucleus than in the overall disk. Being a transitional disk, with a central 
surface brightness in between HSB and LSB spirals, it is not unexpected to find 
that UGC 6879 is a strong IR emitting source compared to the LSB galaxies. Both 
PAH emission and warm dust (24 $\mu$m ) emission show spatial variation along 
the disk. These emissions peak at the central region and diminish toward the edge. 
\subsection{Infrared Diagnostics}
\subsubsection{IR Spectral Energy Distributions}
The flux densities obtained from the 2MASS and {\it IRAS} archives and those 
estimated from the IRAC and MIPS images are given in Table \ref{flux_table}. 
These flux densities are used to construct the observed infrared SEDs of LSB 
galaxies as shown in Fig. \ref{obse_seds}. The open and solid circles represent, 
respectively, 2MASS and IRAC data. The open triangles represent the MIPS data 
whereas the {\it IRAS} upper limits for UGC 6614 and UGC 6879 are shown by filled 
triangles. 
Since Malin 1 was undetected by the MIPS far-IR channels we show  the detection 
limits at 70 and 160 $\mu$m for the total integration time ($\sim$252 sec. and 
$\sim$42 sec. respectively).
We include four SINGS galaxies with different ISM properties for comparison. 
These are NGC 0337 (a normal star forming galaxy), NGC 2798  (a star burst 
galaxy), NGC 2976 (a normal galaxy with nuclear H II region), and NGC 3627 (a 
Seyfert galaxy). They are shown, respectively, by dotted, dashed, dashed-dotted, 
and long dashed lines. The motive is to make a visual comparison between SEDs 
of LSB and HSB galaxies. All flux densities are normalized by the 3.6 $\mu$m 
flux density. 

There are several noticeable features in these SEDs. 
First, the amplitude of the mid-IR and far-IR dust emissions of LSB galaxies are 
lower compared to those of HSB galaxies. An obvious and expected result is that 
the LSB galaxies have less dust content and hence have lower IR emission.  
Second, Malin 1 is deficient in the integrated 8 $\mu$m emission compared to its 
4.5 $\mu$m emission. This is quite opposite for other LSB galaxies and more like 
the SED of an elliptical galaxy (see Fig. 4 in Pahre et al. 2004). 
Third, the 24 $\mu$m emission is suppressed in both Malin 1 and UGC 9024. For 
UGC 6614 it is slightly above the IRAC bands. This feature suggests that the ISM 
of UGC 9024 lacks warm dust emission, and is made mostly of cool dust.
Fourth, the SEDs show a tendency to turnover at relatively longer wavelength, a 
signature that the low density and low surface brightness ISM have low radiation 
intensity. On the other hand, the shape of the SED of UGC 9024 is quite similar 
to those of the representative HSB galaxies. We discuss this in detail in section 
$\S3.3$ 
\subsubsection{Dust Mass}
Dust mass is frequently estimated by fitting the far-IR peak by a single 
temperature modified blackbody function (Hildebrand 1983). However, the 
inability of a single temperature or even two temperature blackbody function 
to fit the observed flux densities suggests that a more sophisticated model 
of the IR SED is needed. The global SED models of Dale et al. (2001) and Dale 
\& Helou (2002; hereafter DH02) provide a robust treatment of the multiple 
grain populations that contribute to the IR emission in a galaxy. This model 
allows a realistic derivation of dust mass since it combines information from 
the full range of heating environments ($\sim$10K-1000K). Previous studies have 
shown that dust mass is underestimated by a factor of $\sim$5-10 for quiescent 
galaxies (i.e. IR cool) if one simply fits the far-IR and sub-millimeter 
continuum data points with a simple single temperature black body instead of 
exploiting information from the full range of the SED.

Figure \ref{glob_seds} shows the fits to the observed SEDs obtained from the 
DH02 model (solid line). The dashed and dotted lines represent, respectively, 
empirical stellar SED (Pahre et al. 2004) and stellar synthesis model prediction 
fitted only to the 2MASS fluxes from Vazquez \& Leitherer (2005). We use the 
model fit (solid line) to estimate $\rm M_d$ and $\mathcal D$ noting that DH02 
model does not provide ideal fits to very extended, low density, diffuse disks 
like Malin 1 and UGC 9024. However, within the measurement and observational 
uncertainties the model fits provide useful insight. A rigorous and detailed 
treatment of infrared SEDs of LSB disks will be presented in a future study. 

The estimated mass is given in Table \ref{basic_table}. We find that the ISM 
of UGC 6614 has the highest amount of dust with dust-to-gas ratio, 
$\mathcal D \sim$0.01. Both Malin 1 and UGC 9024 are $\sim$3 times less dusty 
than UGC 6614. Given that IR emission coming only from the central regions of 
the latter two galaxies, it is not surprising that they show low dust content. 
In a recent study Das et al. (2006) detected CO(1-0) emission localized in a 
specific region on the disk of UGC 6614. They estimated molecular gas mass 
($\rm M_{H_2} \sim 2.8 \times 10^8 M_{\odot}$) which is almost equal to the 
total dust content ($\rm M_d \sim 2.6 \times 10^8$ $M_{\odot}$) that we measure 
distributed over the bulge and disk. The difference between $\mathcal D$ and 
dust-to-(total) gas mass is negligible. 

The (systematic) calibration uncertainty in the observed flux densities and 
the uncertainty in the distance estimates result in a $\sim$30\% uncertainty 
in $\rm M_d$. Additional uncertainty comes from the mass absorption coefficient. 
The different sizes used to measure IR and H I fluxes will also attribute 
additional uncertainty in $\mathcal D$. Besides, considering that the 
long-wavelength end of the SED is poorly constrained the estimate of overall 
dust mass and gas-to-dust ratio is a bit uncertain. All of these errors compound 
to make M$_d$ and $\mathcal D$ uncertain by a factor of $\sim$2 or more.  
\subsubsection{Infrared Luminosity}
DH02 proposed a simple relation to compute total IR luminosity using the MIPS 
bands (see Eq. 4 in DH02). Due to uncertainties in the MIPS flux densities for 
Malin 1 and UGC 9024 we use the empirical relation given by Calzetti et al. 
(2005) to estimate $\rm L_{TIR}$. Calzetti et al. relate flux densities at 8 
$\mu$m and 24 $\mu$m  to derive $L_{TIR}$ for M 51, a normal star forming 
galaxy. 
Estimated total IR luminosities are given in Table \ref{basic_table} with 
estimated uncertainty of $\sim$35\%. We find comparable IR luminosity for both 
Malin 1 and UGC 6614. UGC 9024 is the least luminous because of its suppressed 
24 $\mu$m warm dust emission. In spite of its border line HSB nature, the IR 
output of UGC 6879 resembles a normal quiescent galaxy.
 
We also estimate $\rm L_{TIR}$ using DH02 model fits. We follow Sanders \& Mirabel 
(1996) to define $\rm L_{TIR}$ where flux densities at 12, 25, 60, and 100 $\mu$m 
are obtained from the model SEDs by interpolation. The result is presented in 
Table \ref{basic_table}. Interestingly, these two estimates are within a factor 
of $\sim$2.5 where the model estimates are always higher. The largest difference 
is noted for UGC 9024 which is a result of relatively poor fit to the data.  

Within the uncertainty, the IR energy output of LSB galaxies are smaller by a 
factor of a few compared to their $B$-band luminosities $L_B$. The infrared-to-blue 
ratio, $\rm L_{TIR}/L_{B}$, compares the luminosity processed by dust to that of 
escaping starlight (see Table 2). The ratio ranges from $<$0.01 (in quiescent 
galaxies) to $\sim$100 (in ultra-luminous galaxies). It can be used to 
characterize optical depth of a system composed of dust and stars as well as 
recent ($\sim$100 Myr) SFR to the long term ($\sim$1 Gyr) average rate. 
The ratio is in the range 0.3-0.5 (see Table 2) indicating that the current 
level of star formation is low, a consistent result with previous studies on 
star formation in the LSB ISM (van den Hoek et al. 2000; Burkholder 2001). 
However, there is a potential degeneracy in this parameter and one can only 
make an indirect assessment of the ISM of a galaxy if this degeneracy can be 
lifted. The fact that $\rm L_{TIR}/L_{B}$ is less than unity can arise from two 
different physical conditions. In one hand,  a galaxy may be undergoing intense 
heating by young stars (large $\rm L_B$) but have very little neutral ISM left 
(less IR emission) resulting in low $\rm L_{TIR}/L_{B}$. On the other hand, a 
quiescent galaxy may generate most of its IR emission in H I clouds heated by 
older stellar populations and will display a similarly low $\rm L_{TIR}/L_{B}$ 
(Helou 2000). 
\subsubsection{Star Formation Rate}
A widely used recipe for estimating SFR from IR luminosity is given by Kennicutt 
(1998). However, far-IR luminosity in Kennicutt model is based on {\it IRAS} data. 
Without a proper calibration between {\it IRAS} and MIPS far-IR flux densities the 
uncertainty will loom large in the SFR estimate. We use a new SFR estimator, 
derived recently by Alonso-Herrero et al. (2006; hereafter AH06) using 24 $\mu$m 
flux density. Our estimates are given in Table 2. The error associated with this 
SFR is $\sim$10\%. 

van den Hoek et al. (2000) derived a current SFR ranging $\sim$0.02-0.2 
M$_{\odot}$ yr$^{-1}$ from $I$-band photometry for a sample of LSB galaxies 
generally found in the field. Their estimate agree within a factor of two with 
our results based on infrared data. For Malin 1 and UGC 6614 the infrared SFR 
are $\sim$0.38 M$_{\odot}$ yr$^{-1}$ and $\sim$0.88 M$_{\odot}$ yr$^{-1}$ , 
respectively, whereas it is $\sim$0.01 M$_{\odot}$ yr$^{-1}$ for UGC 9024.
The fact that these galaxies have low dust content indicates that extinction is 
less likely to cause the difference between SFR derived from IR and optical data. 
The higher SFR of UGC 6614 compared to the other two galaxies is consistent with 
the fact that this galaxy has a prominent H$\alpha$ morphology which indicate 
modest level of current star formation. A very low SFR for UGC 9024 implies that 
more light is scattered off from the central disk than being absorbed by the ISM.

The SFR of these three LSB galaxies, as derived from IR data, are thus 
significantly below the rate $\sim$5-10 M$_{\odot}$ yr$^{-1}$ derived for the HSB 
galaxies (Kennicutt 1998) but considerably larger than the rate $\sim$0.001-0.01 
$\rm M_{\odot}$ yr$^{-1}$ observed typically in dwarf irregular galaxies (Hunter 
\& Elmegreen 2004).

The star formation efficiency (SFE), quantified by $\rm L_{TIR}/M_{H I}$, is 
also shown in Table \ref{basic_table}. This measure represents the amount of 
unprocessed gas available to be consumed in subsequent star formation. As 
expected the LSB galaxies have a SFE that is $\leq 1/5$ of the SFE of HSB 
galaxies. 
\subsection{Infrared Colors}
Panels \ref{colo_colo}a and \ref{colo_colo}b show, respectively, mid-IR colors 
and the well known PAH-metallicity connection. In these diagrams we highlight 
two extreme classes (in terms of their IR SEDs) of HSB galaxies: the dwarf 
systems represented by the decimal numerals and the massive elliptical galaxies 
represented by the roman numerals. All of these galaxies are obtained from the 
SINGS sample.  

Panel \ref{colo_colo}a compares PAH emission in various classes of galaxies. 
Metal-rich HSB galaxies preferentially show high ratio in $S_{8\mu m}/S_{24\mu m}$ 
and low ratios of non-stellar 4.5-to-8 $\mu$m dust emission. 
Metal poor (e.g. dwarfs and irregulars) and extremely metal poor HSB galaxies 
such as blue compact dwarfs (BCDs), show the opposite trend and fall within the 
PAH-deficient box (Engelbracht et al. 2005; hereafter E05). Note that E05 only 
looked at star-forming galaxies. ``Red and dead'' ellipticals should have little 
or no PAH emission, yet are nowhere near the PAH-deficient region. So this region 
will not necessarily contain all PAH deficient galaxies. Also note that the result 
of E05 could also arise from selection bias since recent studies have shown that 
dwarf galaxies are not necessarily PAH defficient systems (Rosenberg et al. 2006).

Panel \ref{colo_colo}b, on the other hand, illustrates the PAH-$Z$ connection in 
HSB galaxies. E05 showed that galaxies with low PAH emission have relatively 
unpolluted ISMs. They noted a sharp boundary between galaxy metallicity with and 
without PAH emission, although the trend may have been affected by selection 
bias. We show this trend in panel \ref{colo_colo}b, where the PAH deficient 
galaxies reside inside the dashed region and galaxies with higher metallicites 
avoid two regions in the diagram which are shown by the dotted lines. We are 
interested to see where the LSB galaxies fit in these diagrams.

Panel \ref{colo_colo}a shows that the LSB galaxies are different than the 
PAH-deficient dwarf galaxies in terms of their mid-IR colors. In the color 
space, LSB galaxies occupy a region similar to HSB galaxies and reside 
significantly farther ($> 3\sigma$, along the horizontal axis) from PAH-deficient 
galaxies. While UGC 6614 stays right in the middle of the locus, both Malin 1 and 
UGC 9024 falls in the edge because of the shapes of their SEDs at 8 and 24 $\mu$m.
The mid-IR colors of these galaxies closely resemble those of elliptical galaxies 
which is surprising given that their apparent different star formation histories. 

The LSB galaxies are metal poor with $Z \leq 1/3 Z_{\odot}$ (McGaugh 1994; de 
Bloke \& van der Hulst 1998b; de Naray et a. 2004). McGaugh (1994) provided an 
estimate of oxygen abundance for UGC 6614 but it is highly uncertain. Following 
the general trend shown by the LSB galaxies we assign one-third solar value to 
Malin 1 and a solar value to UGC 6879. The full range and the median values of 
published oxygen abundance are shown for UGC 6614 and UGC 9024.Having very 
limited information it is, therefore, extremely difficult to explore the 
PAH-metallicity connection for these galaxies. We are interested in the question 
whether LSB galaxies being low $Z$ systems, will appear close to the PAH-deficient 
box or will they fall in the region shunned both by HSB dwarfs and HSB spirals 
with extended disks (dotted regions; Fig \ref{colo_colo}b)? While it is tempting 
to give more weight to the latter case, only three data points with large errors 
are insufficient to derive any trend. A larger sample of LSB galaxies with 8 
$\mu$m detections is needed to shed more light on this topic.  

Panel \ref{colo_colo}c shows the connection between mid-IR and far-IR colors 
which basically describe the nature of dust emission at these wavelengths. 
In this panel, the far-IR color of Malin 1 is shown with respect to a far-IR 
flat SED, i.e. same flux density at 70 and 160 $\mu$m. For the other two 
galaxies the diagram shows rather low $S_{70 \mu m}/S_{160 \mu m}$ and high 
$S_{8 \mu m}/S_{24 \mu m}$ ratios. While low far-IR color suggest that they 
are IR cool sources,  the mid-IR color may be linked with the destruction of 
very small grains (but not the PAH molecules) and thus can be used as a 
parameter which can signal evolutionary stages of an ISM. 
A higher value of $S_{8 \mu m}/S_{24 \mu m}$ can mean that $S_{8 \mu m}$ is 
high (large amount of PAH) or that $S_{24 \mu m}$ is low (little or no warm 
dust). Ellipticals are in the latter category (see Pahre et al. 2004) whereas 
LSB systems are in the first category.
It should be noted that the 24 $\mu$m emission is very closely associated 
with H II regions (Helou et al. 2004). Therefore, the lack of emission at this 
wavelength is more likely a deficiency in H II region which is quite consistent 
with the H$\alpha$ images of LSB galaxies (McGaugh et al. 1995; de Naray et al. 
2004).

The vertical arrow in panel \ref{colo_colo}d represents a probable range of 
{\it IRAS} far-IR color for UGC 9024 assuming that {\it Spitzer} far-IR ratio 
is the same as the {\it IRAS} far-IR ratio. The sequence of IR colors can be 
associated with a progression toward greater dust heating intensity and thus 
with a sequence of star formation activity. 
The cool end of the color sequence corresponds to cool diffuse H I medium 
and quiescent molecular clouds, whereas the warm end corresponds to the 
colors of H II regions, starbursts, and galaxies with higher $\rm L_{TIR}/L_B$ 
ratios. Although the IR nature of these three LSB galaxies are explicit in 
this diagram, the interesting feature is that they are not the extreme cases 
in terms of IR SEDs as shown by some of the SINGS galaxies. 

Two primary sources have been proposed to explain the heating of the dust 
which produces IR luminosity in spiral galaxies - massive young (OB) stars 
and associated H II regions (Helou et al. 1985; Devereux \& Young 1990, 1993), 
and non-ionizing A and later stars (Lonsdale \& Helou 1986; Walterbos \& 
Schwering 1987; Bothun et al. 1989). Some authors, however, suggest 
contribution from both sources (Smith 1982; Habing et al. 1984; Rice et al. 
1990; Sauvage \& Thuan 1992; Smith et al. 1994; Devereux et al. 1994).
The 
dominance of the heating source is, therefore, governed by the availability 
of discrete and dense star forming regions in the ISM. While observational 
evidence suggests that the global IR emission from luminous IR spiral galaxies 
provides a measure of high mass SFR (Kennicutt 1998), the diffuse IR emission 
in quiescent galaxies is caused mainly by the thermal radiation of the 
interstellar dust heated by the interstellar radiation field (ISRF) (Jura 1982; 
Mezner et al. 1982; Cox et al. 1986; Jura et al. 1987).

The dust color temperature ($\rm T_d$) deduced for galaxies from the {\it IRAS} 
60 and 100 $\mu$m flux densities are typically 25-40 K assuming emissivity index 
$\beta =2$ (Rahman et al. 2006). This range is similar to the temperature of 
dust in Galactic star-forming regions (Wynn-Williams \& Becklin 1974; Scoville 
\& Good 1989), and considerably greater than the 15-20K temperatures expected 
for dust heated by the ambient ISRF (Jura 1982; Mezner et al. 1982; Cox et al. 
1986; Jura et al. 1987). The color temperature derived from the MIPS 70 and 160 
$\mu$m of LSB galaxies ranges $\sim$17-21 K.

From a statistical study of a large sample of optically selected galaxies, Bothun 
et al. (1989) demonstrated that in the absence of UV radiation, far-IR color ratio 
of $S_{60 \mu m}/S_{100 \mu m} \leq 0.3$ can result from dust which is heated by 
old stars. Galaxies with $S_{60 \mu m}/S_{100 \mu m} \approx  0.3-0.5$ requires a 
steadily increasing proportion of UV heated dust, while galaxies with  
$S_{60 \mu m}/S_{100 \mu m} \geq 0.5$ are entirely dominated by the UV heated dust. 
From panel \ref{colo_colo}d we find that the LSB galaxies have cool effective dust 
temperatures and therefore lack intense heating from massive stars.  

It should be mentioned here that in the IR color analysis we did not subtract the 
stellar contributions from the 8 and 24 $\mu$m measurements to exactly reproduce 
the E05 diagram. Assuming conservatively that the 3.6 $\mu$m emission is coming 
from the stellar photosphere, the stellar contributions obtained from the empirical 
stellar SED of Pahre et al. (2004) derived for early type galaxies are 
$\sim$59\%, $\sim$38\%, and $\sim$19\% at 4.5, 5.8, and 8 $\mu$m, respectively. 
The stellar contribution at 24 $\mu$m emission is $\sim$5\% or less. 
Applying corrections to the corresponding flux densities do not change our results 
since these corrections systematically shift the parameters in the color space. 
Note that we use the same set of numbers for each galaxy ignoring the variation
in amount of stellar contributions from galaxy to galaxy. However, error due to 
this is negligible. To find stellar contributions, preference is given to the 
empirical SED than any stellar synthesis model because of unknown SFHs and poor 
metallicity constraints for LSB galaxies. 

In summary, from the four panels of Fig. \ref{colo_colo} we conclude that:
(a) Malin 1, UGC 6614, and the intermediate HSB/LSB object UGC 6879 have mid-IR 
colors similar to quiescent HSB disk galaxies; UGC 9024 falls in the region of 
HSB elliptical  galaxies in this color plane. (b) There is insufficient data to 
conclude whether LSB galaxies have PAH emission properties significantly different 
from HSB galaxies with comparably low metallicities. Observations of many more 
LSB galaxies are needed to settle this. (c) Available far-IR detections and upper 
limits indicate that LSB galaxies are far-IR cool sources. The dust temperatures 
derived from the MIPS 70 and 160 $\mu$m of LSB galaxies ranges $\rm T_d \sim$17-21 
K, similar to many quiescent HSB spiral and elliptical galaxies.
\subsection{Molecular ISM}
The LSB galaxies are rich in neutral hydrogen (H I). Molecular hydrogen ($\rm H_2$) 
gas inferred from CO emission has been detected in only a handful of such 
galaxies (Matthews \& Gao 2001; O'Neil \& Schinnerer 2004; Matthews et al. 2005; 
Das et al. 2006). The low rate of CO detection has been attributed to an ISM with 
low dust content and a low surface density of neutral gas. 
Dust opacity is crucial for the formation and survival of molecules since it 
provides them necessary shielding from the ISRF. A larger column density is 
needed to self-shield the H$_2$ molecule. A low density and less dusty 
environment exposes H$_2$ to UV photons which can easily dissociate these 
molecules. The low star formation and far-IR cool nature of LSB galaxies implies 
lower energy density of the radiation field and consequently lower dissociation 
of CO and H$_2$ (de Blok \& van der Hulst 1998b).

The deficiency of molecular gas detections in LSB galaxies may point to a 
dynamical condition such as the absence of 
large-scale instability in the disk preventing formation of giant molecular 
clouds (Mihos et al. 1997). Local instabilities may lead to cloud condensation 
resulting in localized star formation which may escape detection by current 
observations (de Blok \& van der Hulst 1998b). Local instability in the disk 
invokes energetic phenomena such as SN explosions and the frequency of such 
occurrence in LSB galaxies is low (Hoeppe et al. 1994; McGaugh 1994). 
The enhanced cooling by molecules is crucial in the onset of instability of 
molecular clouds. Therefore, the effect of less efficient cooling of the ISM 
can also prevent local instabilities. Long cooling time leads to higher cloud 
temperatures and thus makes it difficult for a cold molecular phase to exist 
(Gerritson \& de Blok 1999).

If the low density ISM truly lacks H$_2$ and CO molecules, what other types of 
molecule can exist in this environment? Are they very cold or very warm 
molecules of known types? To which dust components do they belong? Along with 
millimeter wavelength observations, can we use currently available IR data to 
shed light into these questions?
One of the three LSB galaxies, UGC 6614, has been reported to have CO(1-0) 
emission at a certain localized region in the disk (Das et al. 2006), while 
the other two galaxies have not been observed in this wavelength.
On the other hand, the {\it Spitzer} observation of UGC 6614 shows the presence 
of enhanced PAH emission on the bulge and almost entirely along the outer disk. 
This 8 $\mu$m emission, however, is concentrated in the central regions of the 
other two galaxies. 

A possible source of the origin of PAH molecules is the dense, high temperature, 
carbon-rich ([C]/[O] $> 1$) environment of circumstellar envelopes surrounding 
mass-losing 
AGB carbon stars. PAH formation by stars with more normal, oxygen-rich, 
photospheric abundances ([C]/[O] $< 1$) will be negligible because nearly all 
the available carbon is bound up in the CO molecules (Latter 1991). Therefore, 
if the stellar population responsible for the enrichment of the ISM are 
dominantly old carbon-rich stars, the warm PAH molecules will be ubiquitous 
resulting in lower abundance of CO molecules. 
While a detailed investigation is beyond the scope of this study, we believe 
the observed PAH emission and lack of CO emission holds the potential clue 
to probe not only the ISM  but also to better understand the SFHs in LSB galaxies.
\subsection{Mid-IR Photometry of UGC 6614}
The optical spectra of large LSB disks show an unexpected high occurrence of 
low-level active galactic nuclei (AGN) type activity (Spraybarry et al. 1995; 
Schombert 1998). UGC 6614, a optical giant LSB galaxy, is suspected to harbor 
a weak AGN from the optical spectrum (Schombert 1998) and from excess emissions 
at millimeter (Das et al. 2006) and radio wavelengths (Condon et al. 2000). 
Integrated mid-IR photometry provides a robust technique to identify AGN in HSB 
galaxies where AGN tend to be redder than normal star forming galaxies in the 
mid-IR (Lacy et al. 2004; Stern et al. 2005). 

To determine whether one can use a similar technique to detect AGN signatures  
in a LSB bulge, we analyze the IRAC colors ([3.6]-[4.5] vs. [5.8]-[8.0]) of UGC 
6614. The colors are measured for two regions of radius $\sim$1 and $\sim$5.5 
kpc, respectively, encircling the galaxy center (Fig. \ref{agn}; left panel). 
We find that in the color space, along the vertical [3.6]-[4.5] axis, the galaxy 
resides well outside the Stern et al ``AGN box''. Although the lower end of 
the AGN box is within the error bar of this galaxy, its overall mid-IR color 
put it in a region occupied mostly by star forming galaxies suggesting that 
broad band colors may not be an efficient tracer for weak AGNs. 

The contribution from an AGN to the measured IRAC fluxes can be estimated by 
combining image subtraction with assumptions about the nature of the SED coming 
from starlight and AGN emission.  We use Pahre et al. (2004) for stellar flux 
ratios, and a $\nu^{-1}$ power law SED for AGN (Clavel et al. 2000). 
Following the procedure of Howell et al. (2007), a 4.5 $\mu$m image of the 
non-stellar emission was constructed (Fig. \ref{agn}; right panel). 
Unlike the procedure of E05 which simply measures $S_{4.5} - \alpha S_{3.6}$, 
this procedure includes an additional factor to account for the contribution 
of non-stellar emission to the 3.6 $\mu$m image.  
The non-stellar flux density measured this way indicates that, 
within a 12\arcsec \ aperture,  the AGN contributes $\sim$12\% of the 
light at 4.5 $\mu$m and $\sim$6\% at 3.6 $\mu$m.  At 8 $\mu$m, starlight and 
the AGN each contribute $\sim$35\% of the light, with PAHs contributing the 
remaining. Note that the selected aperture includes only the bulge of the 
galaxy, excluding the spiral arm/ring structure.
                                                                        
UGC 6614 illustrates that although [3.6]-[4.5] color can identify strong AGN, 
weaker AGN will not be clearly separated from pure stellar sources. The 
procedure of E05, measuring $S_{4.5 \mu m}-\alpha S_{3.6 \mu m}$, will identify 
regions of non-stellar emission but will not provide quantitative picture of 
stellar emission. Given reasonable assumptions the procedure of Howell et al. 
allows a quantitative decomposition of the stellar and non-stellar flux densities. 
\section{Summary and Conclusions}
The {\it Spitzer} observations of the three optical giant low surface brightness 
galaxies Malin 1, UGC 6614, and UGC 9024 have been examined to study the mid 
and far-IR morphology, spectral energy distributions, and IR color to estimate 
dust mass, dust-to-(atomic) gas mass ratio, total IR luminosity, and star 
formation rate (SFR). We also investigate UGC 6879, which is intermediate 
between HSB and LSB galaxies. 

The 8 $\mu$m images indicate that polycyclic aromatic hydrocarbon (PAH) 
molecules are present in the central regions of all three metal-poor LSB 
galaxies. The diffuse optical disks of Malin 1 and
UGC 9024 remain undetected at mid- and far-infrared wavelengths. The
dustiest of the three LSB galaxies, UGC 6614, has infrared morphology that
varies significantly with wavelength; 160 $\mu$m  (cool) dust
emission is concentrated in two clumps on the NE and NW sides of a distinct
ring seen in the 24 and 8 $\mu$m images (and a broken ring at 70 $\mu$m)
at a radius of $\sim$40\arcsec \ (18 kpc) from the galaxy center.
The 8 and 24 $\mu$m emission is co-spatial with H$\alpha$ emission 
previously observed in the outer ring of UGC 6614. The estimated dust-to-gas 
ratios, from less than $10^{-3}$ to $10^{-2}$, support previous indications 
that LSB galaxies are relatively dust poor compared to HSB galaxies.
The total infrared luminosities are approximately 1/3 to 1/2 the blue band 
luminosities, suggesting that old stellar populations are the primary source 
of dust heating in these LSB objects. The SFR estimated from the infrared 
data ranges $\sim$$\rm 0.01-0.88~M_\odot~yr^{-1}$, consistent with results 
from optical studies. The mid-IR colors of UGC 6614 shows the presence of a 
weak AGN at the central bulge.

\begin{figure}
\epsscale{1.0}
\plotone{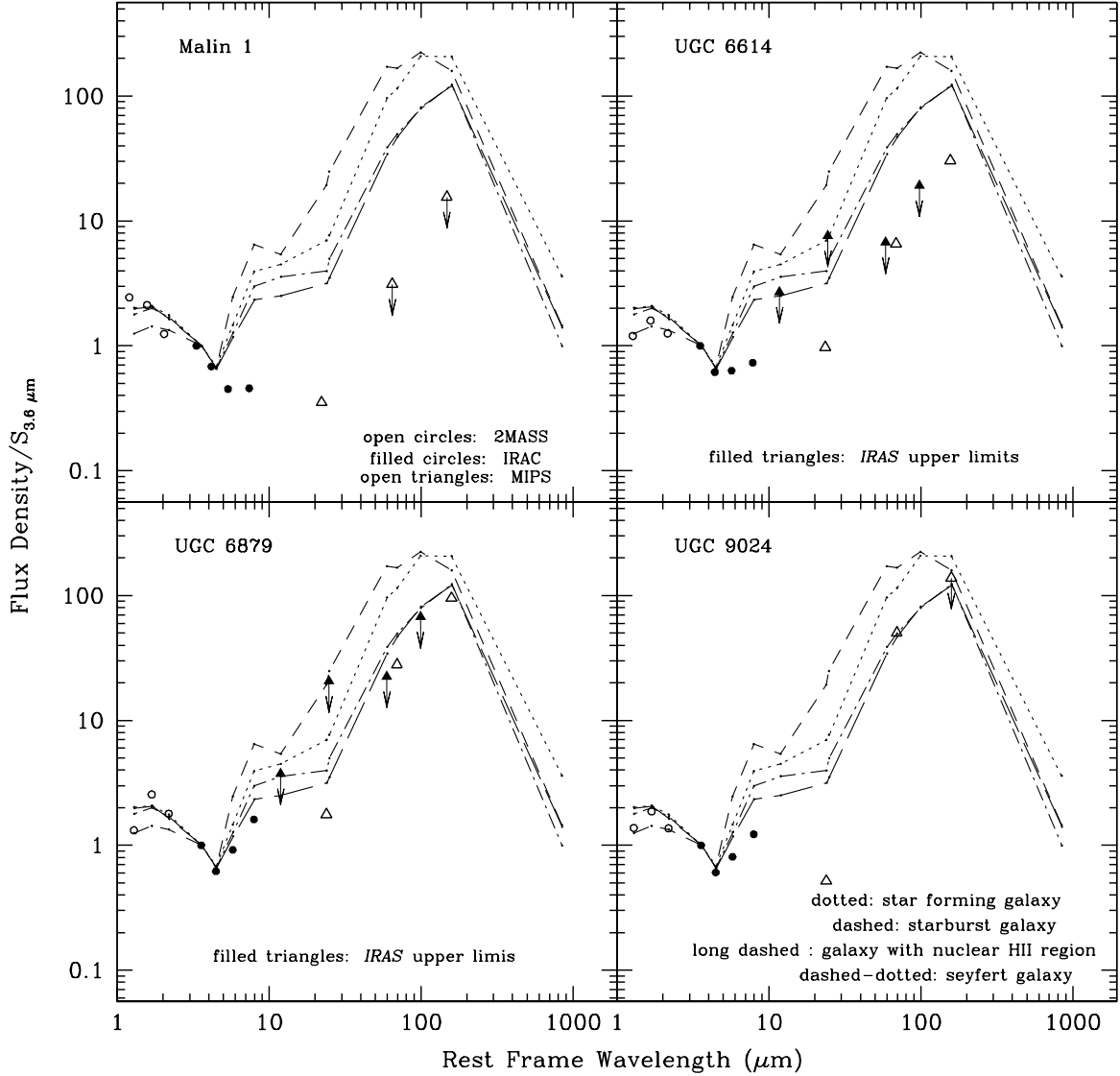}
\figcaption{\small
The observed SEDs of LSB galaxies and UGC 6879 at near, mid, and far-IR 
wavelengths. The 2MASS, IRAC, and MIPS points are shown by the open circles, 
filled circles, and open triangles, respectively. The {\it IRAS} upper limits 
are shown by the filled triangles. For all galaxies the flux densities are 
normalized at 3.6 $\mu$m. Dotted, dashed, dashed-dotted, and long dashed 
lines are used, respectively, to show the SEDs of: 
NGC 0337 (normal star forming galaxy), NGC 2798 (starburst galaxy), NGC 2976 
(galaxy with nuclear H II region), and NGC 3627 (seyfert II galaxy). Malin 1 
was undetected by the MIPS far-IR channels and hence the detection limits are 
shown for the total integration time ($\sim$252 sec. at 70 $\mu$m and $\sim$42 
sec. at 160 $\mu$m). \label{obse_seds}}
\end{figure}

\begin{figure}
\epsscale{1.0}
\plotone{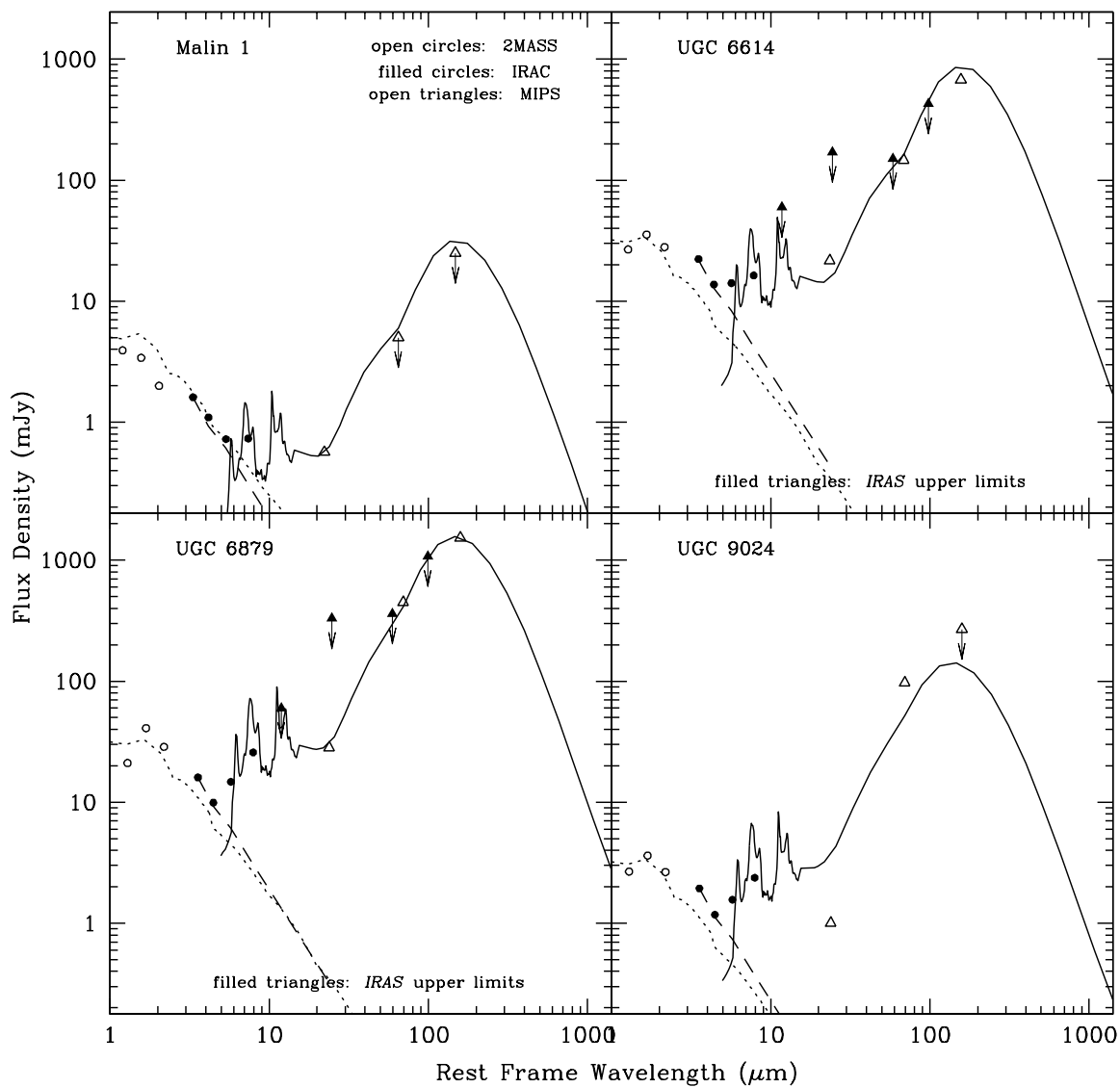}
\figcaption{\small
The observed SEDs of LSB galaxies and UGC 6879 using the DH02 model. The symbol 
styles are similar to Fig. \ref{obse_seds}. The dashed represents the empirical 
stellar SED of Pahre et al. (2004). The dotted lines represents the stellar 
synthesis model prediction from Vazquez \& Leitherer (2005) fitted only to the 
2MASS fluxes. The dust mass is estimated using the fitted SEDs (solid line).
\label{glob_seds}}
\end{figure}

\begin{figure}
\epsscale{0.75}
\plotone{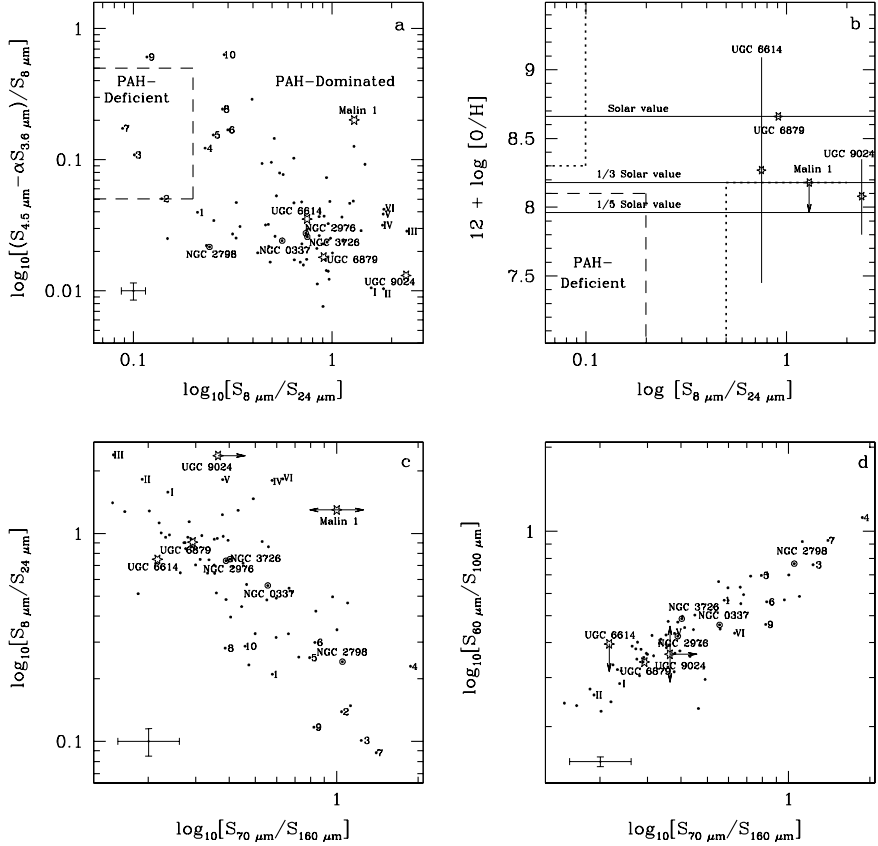}
\figcaption{\small
The mid and far-IR color-color diagrams highlighting the LSB spirals (shown by 
open stars) with respect to different classes of HSB galaxies (shown by black 
dots). The SINGS (Kennicutt et al. 2003 ) sample galaxies are taken as the 
representative of local HSB galaxies. This sample contains various types of 
galaxies such as dwarfs, normal starforming, starburst, Syfert I, Seyfert II 
and ellipticals. 
In all panels galaxies represented by the decimal numerals are dwarfs systems 
whereas those shown by the roman numerals are ellipticals. The rest of the 
points (black dots) represent other population types where one galaxy from each 
population are shown by the open circle: NGC 0337 (normal star forming galaxy), 
NGC 2798 (starburst galaxy), NGC 2976 (galaxy with nuclear H II region), and 
NGC 3627 (Seyfert II galaxy).
In panel (a) $\alpha \approx 0.58$ is the stellar contribution at 4.5 $\mu$m 
estimated from the empirically derived stellar SED of Pahre et al. (2004). 
The regions covered by the dotted boxes (panel b) are from a similar diagram of 
Eangelbracht et al. (2005). No HSB galaxies (dwarfs or extended disks) occupy 
these regions. Far-IR color of Malin 1 (panel c) is shown with respect to a flat 
far-infrared SED. The vertical arrow in panel (d) represents a probable range of 
{\it IRAS} color for UGC 9024. A few galaxies shown by numerals do not appear in 
panel (d) because of a lack of {\it IRAS} data. Representative error bars based 
on calibration uncertainty are shown. \label{colo_colo}}
\end{figure}

\begin{figure*}[ht]
\begin{center}
\includegraphics[width=2.45in]{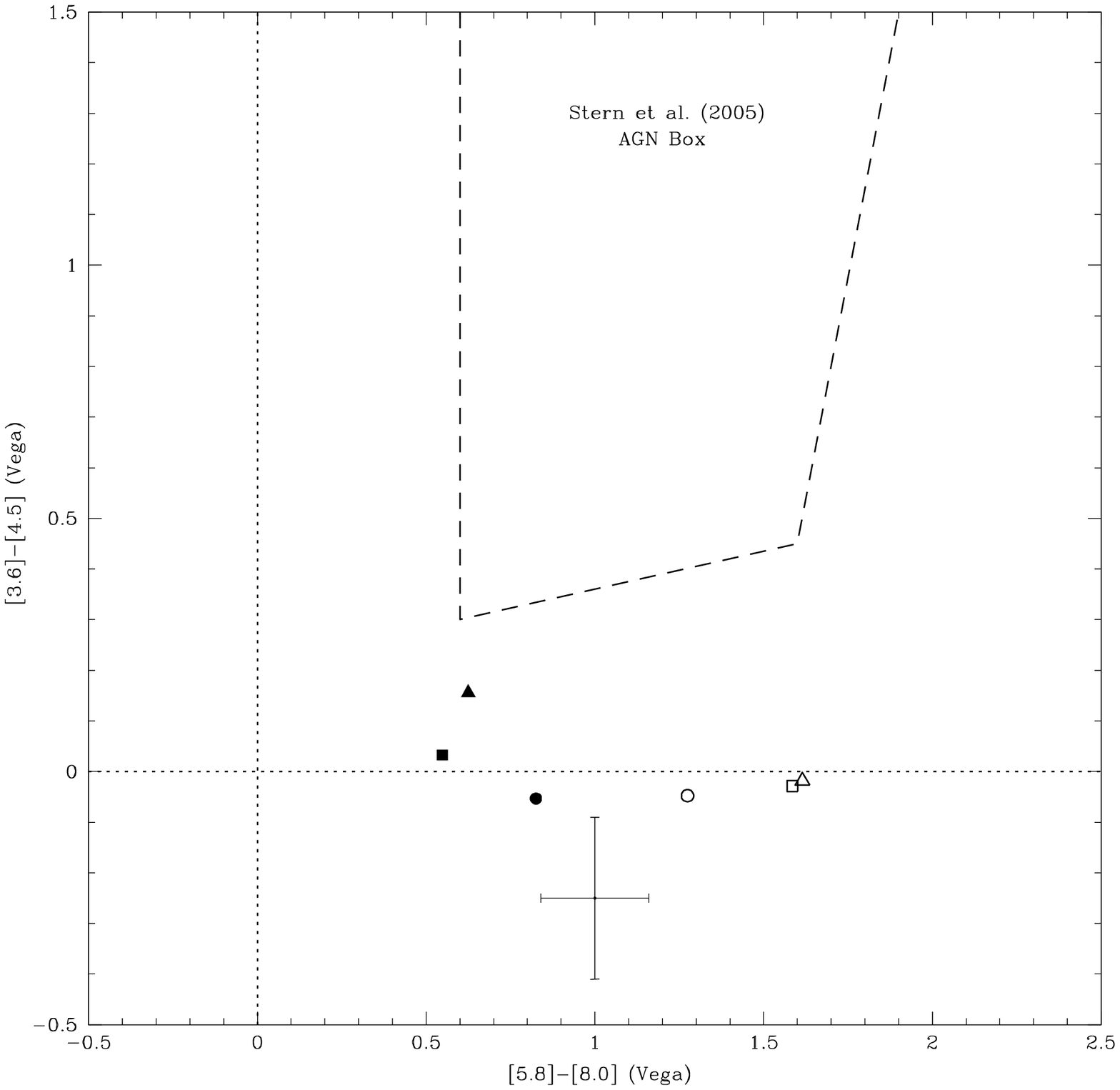} 
\includegraphics[width=2.45in]{f10.eps}
\end{center}
\caption{\small 
Left panel: The IRAC colors for UGC 6614 as shown by solid symbols. The triangle 
(square) represents the color measured within a central region of radius of 
2\arcsec \ (12\arcsec); \ this corresponds to a physical radius 
of $\sim$1 kpc ($\sim$5.5 kpc). The circle represents integrated flux from the 
entire galaxy (Table \ref{flux_table} and \ref{basic_table}). 
In this diagram, stars occupy a locus around (0,0) with various exceptions 
(Stern et al. 2005). Star forming galaxies mostly lie along the the horizontal 
axis depending on the amount of 8 $\mu$m PAH emission. UGC 6879 is shown by open 
symbols to highlight the behavior of non-AGN type galaxies. 
A typical error bar is shown at the bottom. Right panel: an image 
(radius 12\arcsec) of non-stellar emission at 4.5 $\mu$m after taking 
into account the contribution of similar emission at 3.6 $\mu$m image. See text 
for detail. \label{agn}}
\end{figure*}

Questions can be raised such as what is the most viable reason for these LSB 
galaxies to have $\rm L_{TIR}/L_{B} < 1$? To answer this question we first 
note that observables such as stellar populations and SED shapes can be used 
to break the degeneracy in infrared-to-blue ratio (Helou 2000). That LSB 
galaxies have low infrared-to-blue luminosities, stellar populations spanning 
a wide range of mean ages, are not dominated by OB stars (McGaugh 1994), have 
less dust than the HSB galaxies (see Fig. \ref{obse_seds}), and are IR cool 
sources (see Fig. \ref{colo_colo}) suggest a composite scenario. The LSB disks 
are less dusty and the older stellar populations are the primary source of the 
IR emission from their ISMs.

The presence of PAH emission in these three galaxies indicates that the ISM 
of the region contributing to the emission in these galaxies have significant 
amount of carbon enrichment over cosmic time and the ISRF in the ISM must have 
been significantly weak and thus it is unable to reduce the strength of PAH 
emission. In other words, the small grains are more exposed to the ISRF so that 
their destruction rate is larger than for PAH molecules. 

The detection of mid and far-IR emission from a larger sample will be crucial 
to understand the properties of ISM in LSB galaxies and probing their star 
formation histories. This will have a significant effect on analytical modeling 
of galaxy formation and evolution, the role of different galaxy populations in 
observed number counts and possibly metallicity effects in the observed number 
counts. Whether star formation in LSB disks occurred in a continuous fashion 
but with a low rate, or in an exponentially decreasing rate, or as sporadic 
bursts with quiescent periods in between is still a matter of debate. Since 
each type of formation history will lead to a stellar population 
that could be traced by optical photometry (i.e. blue or red), the formation 
scenario must follow the route where the ISM would be in a state having 
significant carbon enrichment with substantial amount of dust. The constraints 
coming from {\it SST} such as mid-IR 8 $\mu$m emission and moderate dust mass 
could be used as probes to understand the nature of LSB spirals.  

Metal poor HSB objects such as blue compact dwarf galaxies are PAH deficient 
systems (E05; Wu et al. 2006). Their SEDs are markedly different in the 
$\sim$5-15 $\mu$m wavelength range compared to metal rich HSB galaxies. The 
LSB galaxies are metal poor but have substantial PAH emission. Although these 
galaxies are not the extreme cases in metal deficiency such as BCDs, they fill 
an interesting niche among local populations, distinct from HSB dwarfs and 
from HSB regular galaxies. They may also represent a significant fraction of 
the galaxy population at earlier epochs (Zackrisson et al. 2005), and therefore, 
may have important implication in the interpretation of galaxy number counts 
in the infrared/submillimeter as well as in the visible and near-IR wavelengths.  
 
Previous analytical studies suggest that metal abundances have profound 
implications on galaxy number counts observed at 24, 70, and 160 $\mu$m  
(Lagache et al. 2003; Dale et al. 2005). To date, in analytical models 
metallicity effects have been incorporated in an ad hoc manner by artificially 
manipulating SEDs of HSB galaxies in the wavelength range mentioned above 
(Lagache et al. 2003). When template SEDs of many nearby metal-poor LSB 
galaxies become available, one can incorporate these in galaxy evolution models 
as an independent class along with various other classes such as normal star 
forming, starburst, luminous and ultra-luminous, and AGNs to understand the 
observed galaxy number counts and the origin of the IR background.  
\acknowledgments
The anonymous referee is thanked for constructive comments and suggestions. 
We happily thank D. Dale for his model fits. We also thank Y. Wu, B. R. Brandl, 
J. R. Houck for helpful communications. We acknowledge useful discussions from 
A. Blain, G. D. Bothun, and S. S. McGaugh on LSB galaxy population. One of us 
(NR) gratefully acknowledges the support of a Research Associateship 
administered by Oak Ridge Associated Universities (ORAU) during this research.
This research has made use of the NASA/IPAC Extragalactic Database (NED) which 
is operated by the Jet Propulsion Laboratory, California Institute of Technology, 
USA under contract with the National Aeronautics and Space Administration, and 
the LEDA database in France. This study is based on observations made with the 
{\it Spitzer} Space Telescope, which is operated by the Jet Propulsion 
Laboratory,  California Institute of Technology under NASA contract 1407.
This study has made use of data products from the Two Micron All Sky Survey, 
which is a joint project of the University of Massachusetts and IPAC/Caltech, 
funded by NASA and the National Science Foundation. This study also acknowledges 
use of data products from Solan Digital Sky Survey.

{}
\begin{deluxetable}{cccccccccc}
\tabletypesize{\small}
\tablewidth{0pc}
\setlength{\tabcolsep}{0.04 in} 
\tablecolumns{6}
\tablecaption{Flux densities from 2MASS, {\it IRAS}, and {\it Spitzer}}
\tablehead{ 
\multicolumn{1}{c}{Source}            & 
\multicolumn{1}{c}{$\lambda (\mu m)$} &
\multicolumn{2}{c}{Malin 1}           &
\multicolumn{2}{c}{UGC 6614}          &
\multicolumn{2}{c}{UGC 6879}          &
\multicolumn{2}{c}{UGC 9024}           \\\cline{3-10}
\colhead{}            &\colhead{}  
&\colhead{$S_{\nu}$}  &\colhead{$r$}
&\colhead{$S_{\nu}$}  &\colhead{$r$}
&\colhead{$S_{\nu}$}  &\colhead{$r$}
&\colhead{$S_{\nu}$}  &\colhead{$r$} \\
\colhead{}        &\colhead{}  
&\colhead{(mJy)}  &\colhead{(arcsec)}
&\colhead{(mJy)}  &\colhead{(arcsec)}
&\colhead{(mJy)}  &\colhead{(arcsec)}
&\colhead{(mJy)}  &\colhead{(arcsec)} 
}
\startdata 
2MASS     &1.3  &3.93   &$10$ &26.80   &$26$ &21.10  &$63$ &2.67 &$13$ \\
2MASS     &1.7  &3.41   &$10$ &35.50   &$26$ &40.90  &$63$ &3.62 &$13$ \\
2MASS     &2.2  &2.00   &$10$ &28.00   &$26$ &28.70  &$63$ &2.65 &$13$  \\ 
\hline
IRAC      &3.6  &1.61   &$12$ &22.31   &$60$ &16.03  &$75$ &1.94  &$24$ \\ 
IRAC      &4.5  &1.10   &$12$ &13.74   &$60$  &9.93  &$75$ &1.18  &$24$ \\
IRAC      &5.8  &0.73   &$12$ &14.11   &$60$ &14.73  &$75$ &1.57  &$24$ \\ 
IRAC      &8    &0.74   &$12$ &16.33   &$60$ &25.78  &$75$ &2.38  &$24$ \\   
\hline
{\it IRAS} &12   &--     &--        &$<$60   &--        &$<$60  &--        &--    &--   
\\
MIPS      &24   &0.57   &$30$       &22      &$60$ &28.35  &$75$ &1.00  &$30$ \\
{\it IRAS} &25   &--     &--        &$<$170  &--        &$<$330 &--        &--    &--  
\\
{\it IRAS} &60   &--     &--        &$<$150  &--        &$<$360 &--        &--    &--  
\\
MIPS      &70   &$<$5   &--        &147     &$60$ &447    &$75$ &98    &$60$ \\
{\it IRAS} &100  &--     &--        &$<$430  &--        &$<$1080 &--       &--    &--  
\\
MIPS      &160  &$<$25  &--        &677     &$60$ &1529 &$75$  &$<$268 &$60$\\ \hline
\enddata \\  
\tablecomments{The IRAC flux densities are aperture corrected. Malin 1 was 
undetected by the MIPS far-IR channels and hence the detection limits at 70 
and 160 $\mu$m are shown for the total integration time: $\sim$252 sec. at 
70 $\mu$m and $\sim$42 sec. at 160 $\mu$m. In general, $r$ represents the 
radius of a circular aperture which is used to estimate the flux density. 
However, the 2MASS flux densities are given for ``total'' elliptical aperture 
radius in the archive 
(http://irsa.ipac.caltech.edu/applications/2MASS/PubGalPS) where elliptical 
radius represents the semi-major axis. UGC 6879 is edge-on and for this galaxy 
we use ``effective radius'', defined by $r = \sqrt{a b}$, where $a$ and $b$ are 
semi-major and minor axes.} 
\label{flux_table}
\end{deluxetable}
\begin{deluxetable}{cccccc}
\tabletypesize{\small}
\tablewidth{0pc}
\setlength{\tabcolsep}{0.14 in} 
\tablecolumns{6}
\tablecaption{Basic Properties of LSB galaxies}
\tablehead
{ 
\colhead{Property}  &\colhead{Malin 1}  &\colhead{UGC 6614}   
&\colhead{UGC $6879^{\ast}$} &\colhead{UGC 9024}  &\colhead{Reference}
}
\startdata
Type                    &S?          &SAa        &SABc      &SAab       &1,2    \\
redshift, $z$           &0.0834      &0.0237     &0.009     &0.0088     &1,2    \\
D$_A$ (Mpc)             &320.47      &93.51      &35.87     &34.86      &1,2    \\ 
inclination, $i$        &16.1$^{o}$  &34$^{o}$   &90$^{o}$  &32.2$^{o}$ &1,2    \\ 
D$_{25}$ (arcsec)       &16.15       &75.54      &87.75     &104.26     &1,2    \\
D (kpc)                 &25.09       &34.24      &15.25     &17.62      &1,2    \\
$m_B$ (mag)             &17.92       &14.12      &13.40     &17.11      &1,2    \\
$M_B$ (mag)             &-19.83      &-20.70     &-19.52    &-15.64     &1,2    \\
$\mu_{B,0}$ (mag arcsec$^{-2}$)  &26.50  &24.50  &20.40     &24.71      &4,5,6  \\
$\rm h_{r,R}$ (kpc)     &73.3        &16.00      &2.50      &7.47       &4,5    \\
$Y_o$                         &$<$8.66   &7.35-9.16  &--     &7.98-8.35  &3,9   \\
$\log(\rm L_B/L_{\odot})$     &9.94      &10.29      &9.83   &8.21       &1,2,5  \\
$\log(\rm M_{H I}/M_{\odot})$ &10.66     &10.42      &9.08      &9.40       &5,7,8  \\
$\log(\rm M_{H_2}/M_{\odot})$ &--        &8.45       &--        &--         &10    \\
$\log(\rm M_d/M_{\odot})$     &$<$8.14   &8.42       &7.81      &$<$6.94    &11    \\ 
$\mathcal D$                  &$<$0.003    &0.01       &0.053     &$<$0.004  &11    \\ 
$\log(\rm L_{TIR}/L_{\odot})$ &9.49        &9.72       &9.05      &7.90      &11    \\
$\log(\rm L_{TIR}^{\dag}/L_{\odot})$ &$<$9.50  &9.80    &9.28      &$<$8.30  &11    \\
$\rm L_{TIR}/L_B$           &0.35        &0.27       &0.17      &0.49       &11    \\
SFR ($\rm M_{\odot}~yr^{-1}$) &0.38  &0.88       &0.20      &0.01       &11    \\ 
$\rm L_{TIR}/M_{H I} \ (L_{\odot}/M_{\odot})$  &0.07  &0.20  &0.93 &0.03  &11  \\ \hline
\enddata \\
\tablecomments{\small
$^{\ast}$UGC 6879 is not strictly a LSB galaxy. Its central surface brightness 
$\mu_{B,0}$ is intermediate between LSB and HSB galaxies.  
We assume $\Omega_M = 0.3$, $\Omega_{\Lambda} = 0.7$, and $H_0 = 75$ km 
sec$^{-1}$ Mpc$^{-1}$ to be consistent with the literature. 
Notation: 
D$_A$ is the angular diameter distance; 
D$_{25}$ is the $B$-band 25 mag arcsec$^{-2}$ isophotal diameter; D is the
physical diameter; 
$m_B$ and $M_B$, respectively, are $B$-band apparent and absolute magnitude 
corrected for Galactic extinction, internal extinction, and $K$-correction; 
$\mu_{B,0}$ is the $B$-band central surface brightness; disk scale length 
$\rm h_{r,R}$ is in kpc; 
the solar value of the oxygen abundance $Y_o = 12+\log[O/H] \sim 8.66$ from 
Asplund et al. (2004); 
$\rm M_{H I}$ is neutral hydrogen mass;  $\rm M_{H_2}$ is molecular hydrogen mass;  
$\rm M_d$ is dust mass; the dust-to-gas ratio is $\mathcal D = \rm M_d/M_{H I}$;
total IR luminosity $\rm L_{TIR}$ is estimated from Calzetti et al. (2005); 
total IR luminosity $\rm L_{TIR}^{\dag}$ is estimated from the DH02 model fit;
star formation rate SFR is calculated from Alfonso-Herrero et al. (2006). \\
References: 
1) LEDA; 
2) NED; 
3) McGaugh 1994; 
4) Sprayberry et al. 1995; 
5) Impey et al. 1996; 
6) McGaugh \& de Blok 1998; 
7) Matthews et al. 2001;
8) Sauty et al. 2003; 
9) de Naray et al. 2004;
10) Das et al. 2006; 
11) This study.}
\label{basic_table}
\end{deluxetable}


\begin{thebibliography}{}
\bibitem[Alfonso-Herrero et al. 2006]{}
Alfonso-Herrero, A., Reike, G. H., Reike, M. J., Colina, L., 
Perez-Gonzalez, P. G., \& Ryder, S. D. 2006, ApJ, 650, 835

\bibitem[Asplund et al. 2004]{}
Asplund, M., Grevesse, N., Sauval, A. J., Allende Prieto, C., 
\& Kiselman, D. 2004, A\&A, 417, 751

\bibitem[Barth 2007]{}
Barth, A. J., 2007, AJ, 133, 1085

\bibitem[Bell et al. 2000]{}
Bell, E. F., Barnaby, D., Bower, R. G., et al. 2000, MNRAS, 312, 
470 

\bibitem[Bergvall et al. 1999]{}
Bergvall, N., R\~{o}nnback, J., Masegosa, J., \& \"{O}stlin, G. 
1999, A\&A, 314, 679

\bibitem[Bernard et al. 1994]{}
Bernard, J. P., Boulanger, F., D{e}sert, F. X., Giard, M., 
Helou, G., \& Puget, J. L. 1994, A\&A, 291, L5 

\bibitem[Bothun et al. 1997]{}
Bothun, G. D., Impey, C. D., McGaugh, S. S. 1997, PASP, 109, 745 

\bibitem[Bothun et al. 1989]{}
Bothun, G. D., Lonsdale, C. J., \& Rice, W. 1989, ApJ, 341, 129

\bibitem[Burkholder et al. 2001]{}
Burkholder, V., Impey, C., \& Sprayberry, D. 2001, ApJ, 122, 2318

\bibitem[Calzetti et al. 2005]{}
Calzetti, D., Kennicutt, R. C., Jr., Bianchi, L., et al. 2005, 
ApJ, 633, 871

\bibitem[Clavel et al. 2000]{}
Clavel, J., Schulz, B., Altieri, B., et al. 2000, A\&A, 357, 839

\bibitem[Cox et al. 1986]{}
Cox, P., Krugel, E., \& Mezgner, P. 1986, A\&A, 155, 380

\bibitem[Dale et al. 2001]{}
Dale, D., Helou, G., Contursi, A., Silbermann, N. A., \& 
Kolhatkar, S. 2001, 549, 215

\bibitem[Dale \& Helou 2002]{}
Dale, D., \& Helou, G. 2002, 576, 159 (DH02)

\bibitem[Dale et al. 2005]{}
Dale, D., Smith, J. D. T., Armus, L., et al. 2005, ApJ, 646, 161

\bibitem[Das et al. 2006]{}
Das, M., O'Neil, K., Vogel S. N., \& McGaugh, S. S. 2006, ApJ, 
651, 853

\bibitem[de Blok \& van der Hulst 1998a]{}
de Blok, W. J. G., \& van der Hulst, J. M. 1998a, A\&A, 335, 421

\bibitem[de Blok \& van der Hulst 1998b]{}
de Blok, W. J. G., \& van der Hulst, J. M. 1998b, A\&A, 336, 49

\bibitem[de Naray et al. 2004]{}
de Naray R. K., McGaugh S. S., \& de Blok, W. J. G. 2004, MNRAS, 
355, 887

\bibitem[Desert et al. 1990]{}
Desert, F. X., Boulanger, F., \& Pugetet, J. L. 1990, AA, 237, 215

\bibitem[Devereux \& Young 1990]{}
Devereux, N., \& Young, J. 1990, ApJ, 359, 42

\bibitem[Devereux \& Young 1993]{}
Devereux, N., \& Young, J. 1993, AJ, 106, 948

\bibitem[Devereux et al. 1994]{}
Devereux, N., Price, R., Wells, L., \& Duric, N. 1994, AJ, 108, 
1667

\bibitem[Engelbracht et al. 2005]{}
Engelbracht, C. W., Gordon, K. D., Rieke, G. H., Werner, M. W., 
Dale, D. A., \& Latter, W. B. 2005, ApJ, 628, L29 (E05)

\bibitem[Fazio et al. 2004]{}
Fazio, G. G., Hora, J. L., Allen L. E., et al. 2004, ApJS, 154, 
10

\bibitem[Freeman 1970]{}
Freeman, K. C. 1970, ApJ, 160, 811

\bibitem[Gerritsen \& de Blok 1999]{}
Gerritsen, J. P. E., \& de Blok, W. J. G. 1999, A\&A, 342, 655

\bibitem[Gautier 1986]{}
Gautier, T. N. III. 1986, in Light on Dark Matter, 
ed. F. P. Israel (Dordrecht: Reide), p. 49 

\bibitem[Habing et al. 1984]{}
Habing, H. J., Miley, G., Young, E., et al. 1984, ApJ, 278, L59

\bibitem[Helou et al. 1985]{}
Helou, G., Soifer, B. T., \& Rowan-Robinson, M. 1985, 305, L15

\bibitem[Helou  2000]{}
Helou, G. 2000, in Les Houches Summer School, Vol 70, 
Infrared Space Astronomy: Today and Tomorrow, 
ed. F. Casoli, J. Lequex, \& F. David (Springer, 
New York), p. 337  

\bibitem[Helou et al. 2000]{}
Helou, G., Lu, N. Y., Werner M. W., Malhotra, S., 
\& Silbermann, N. 2000, ApJ, 532L, 21 

\bibitem[Helou et al. 2004]{}
Helou, G., Roussel, H., Appleton, P., et al. 2004, ApJS, 154, 
253

\bibitem[Hildebrand 1983]{}
Hildebrand, R. H. 1983, QJRAS, 24, 267 

\bibitem[Hinz et al. 2006]{}
Hinz, J. L., Misselt, K., Rieke, M. J., Rieke G. H., Smith, 
P. S., Blaylock, M., \& Gordon, K. D. 2006, ApJ, 651, 874
 

\bibitem[Hoeppe et al. 1994]{}
Hoeppe, G., Brinks, E., Klein, U. et al. 1994, AJ, 108, 446



\bibitem[Howell et al. 2007]{}
Howell, J. H., Mazzarella, J. M., Chan, B., et al. 2007, in preparation

\bibitem[Hunt et al. 2002]{}
Hunt, L. K., Gio vandani, C., \& Helou, G. 2002, A\&A, 394, 
873


\bibitem[Hunter \& Elmegreen 2004]{}
Hunter, D. A., \& Elmegreen, B. G. 2004, AJ, 128, 2870

\bibitem[Impey \& Bothun 1997]{}
Impey, C. D., \& Bothun, G. D. 1997, ARA\&A, 35, 267 

\bibitem[Impey et al. 1996]{}
Impey, C. D., Sprayberry, D., Irwin, M. J., \& Bothun, G. D. 
1996, ApJS, 105, 209 

\bibitem[Ishida 2004]{}
Ishida, C. M. 2004, Ph.D. Thesis, University of Hawaii 

\bibitem[Jarrett et al. 2000]{}
Jarrett, T. H., Chester, T., Cutri, R., Schneider, S., 
Skrutskie, M., \& Huchra, J. P. 2000, AJ, 119, 2498

\bibitem[Jura 1982]{}
Jura, M. J. 1982, ApJ, 254, 70

\bibitem[Jura et al. 1987]{}
Jura, M. J., Kim, D. W., Knapp, G., \& Guhathakurta, P. 1987, 
ApJ, 312, L11

\bibitem[Kennicutt et al. 2003]{}
Kennicutt, R. C., Jr., Armus, L., Bendo, G., et al. 2003, 
PASP, 115, 928

\bibitem[Kennicutt 1998]{}
Kennicutt, R. C., Jr. 1998, ApJ, 498, 541

\bibitem[Kessler et al. 1996]{}
Kessler, M. F., Steinz, J. A., Anderegg, M. E., et al. 1996, 
A\&A, 315, L27

\bibitem[Lacy et al. 2004]{}
Lacy, M., Storrie-Lombardi, L. J., Sajina, A., et al. 2004, 
ApJS, 154, 166 

\bibitem[Latter 1991]{}
Latter, W. B. 1991, ApJ, 377, 187

\bibitem[Lonsdale \& Helou 1987]{}
Lonsdale, C. J. \& Helou, G. 1987, ApJ, 314, 513

\bibitem[Lu  et al. 2003]{} 
Lu, N., Helou, G., Warner, M. W., et al. 2003, ApJ, 588, L199 

\bibitem[Matthews et al. 1999]{}
Matthews, L. D., Gallagher, J. S., van Driel, W. 1999, 
AJ, 118, 2751

\bibitem[Matthews et al. 2005]{}
Matthews, L. D., Gao, Y., Uson, J. M., \& Combes, F. 2005, 
ApJ, 129, 1849 

\bibitem[McGaugh 1994]{}
McGaugh, S. S. 1994, ApJ, 426, 135 

\bibitem[McGaugh \& Blok 1998]{}
McGaugh, S. S., \& de Blok, W. J. G. 1998, ApJ, 499, 41

\bibitem[McGaugh et al. 1995a]{}
McGaugh, S. S., Schombet, J., \& Bothun, G. D. 1995a, ApJ, 
109, 2019 

\bibitem[McGaugh et al. 1995b]{}
McGaugh, S. S., Bothun, G. D., \& Schombet, J. 1995b, ApJ, 
110, 573

\bibitem[Mihos et al. 1997]{}
Mihos, J. C., McGaugh, S. S., \& de Blok, W. J. G. 1997, ApJ, 
477, L79

\bibitem[Mezger et al. 1982]{}
Mezger, P., Mathis, J., \& Panagia, N. 1982, A\&A, 105, 372 

\bibitem[Mould et al. 2000]{}
Mould, J. R., Huchra, J. P., Freedman, W. L., et al. 2000, 
ApJ, 529, 786

\bibitem[Neugebauer et al. 1984]{}
Neugebauer, G., Habing, H. J., van Duinan, R., et al. 1984, 
ApJL, 278, L1

\bibitem[O'Neil et al. 1997]{} 
O'Neil, K., Bothun, G. D., Schombert, J., Cornellm M. E., \& 
Impey, C. D. 1997, AJ, 114, 2448

\bibitem[O'Neil et al. 1998]{} 
O'Neil, K., Bothun, G. D., Impey, C. D., \& McGaugh, S. S. 
1998, AJ, 116, 657

\bibitem[O'Neil \& Schinnerer 2004]{} 
O'Neil, K., \& Schinnerer, E. 2004, ApJL, 615, 109 

\bibitem[O'Neil et al. 2004]{}
O'Neil, K., Bothun, G. D., van Driel, W., Monnier Ragaigne, D. 
2004, A\&A, 428, 823

\bibitem[Pahre et al. 2004]{}
Pahre, M. A., Ashby, M. L. N., Fazio, G. G., \& Wilner, 
S. P. 2004, ApJS, 154, 229

\bibitem[Pickering \& van der Hulst 1999]{}
Pickering, T. E., \& van der Hulst, T. 1999, Bulletin of the 
American Astronomical Society, Vol. 31, p. 1525

\bibitem[Puget \& Leger 1989]{} 
Puget, J. L., \& Leger, A. 1989, ARA\&A, 27, 161 

\bibitem[Rahman et al. 2006]{}
Rahman, N., Helou, G. \& Mazzarella, J. M. 2006, ApJ, 653, 
1068

\bibitem[Reach et al. 2005]{}
Reach, W. T., Megeath S. T., Cohen, M., et al. 2005, PASP, 
117, 978 

\bibitem[Rice et al. 1990]{}
Rice, W., Boulanger, F., Viallefond, F., Soifer, B. T., \& 
Freedman, W. L. 1990, ApJ, 358, 418

\bibitem[Richer \& McCall 1995]{}
Richer, M. G., \& McCall, M. L. 1995, ApJ, 445, 642

\bibitem[Rieke et al. 2004]{}
Rieke, G. H., Young, E. T., Engelbracht C. W., et al. 2004, 
ApJS, 154, 25

\bibitem[Rosenberg et al. 2006]{}
Rosenberg, J. L., Ashby, M. L. N., Salzer, J. J., \& Huang, 
J.-S., 2006, ApJ, 636, 742

\bibitem[Sanders \& Mirabel 1996]{}
Sanders, D. B. \& Mirabel, I. F. 1996, ARA\&A, 34, 749 

\bibitem[Sauty et al. 2003]{}
Sauty, S., Casoli, F., Boselli, A., et al. 2003, A\&A, 411, 
381 

\bibitem[Sauvage \& Thuan 1992]{}
Sauvage, M., \& Thuan, T. 1992, ApJ, 396, L69

\bibitem[Schombert 1998]{}
Schombert, J. 1998, ApJ, 116, 1650

\bibitem[Schombert et al. 1990]{}
Schombert, J., Bothun, G. D., Impey, C. D. \& Mundy L. G.  
1990, AJ, 100, 1523

\bibitem[Schombert et al. 1992]{}
Schombert, J., Bothun, G. D., Schneider S. E., \& McGaugh 
S. S. 1992, AJ, 103, 1107

\bibitem[Scoville \& Good 1989]{}
Scoville, N. Z., \& Good, J. C. 1989, ApJ, 339, 149

\bibitem[Smith 1982]{}
Smith, J. 1982, ApJ, 261, 463

\bibitem[Smith et al. 1994]{}
Smith, B., Harvey, P., Colome, C., et al. 1994, ApJ, 425, 1994

\bibitem[Sprayberry et al. 1995]{}
Sprayberry, D., Impey C. D., Bothun, G. D., \& Irwin, M. J. 
1995, ApJ, 109, 558

\bibitem[Sprayberry et al. 1997]{}
Sprayberry, D., Impey C. D., Irwin, M. J., \& Bothun, G. D. 
1997, ApJ, 482, 104

\bibitem[Stern et al. 2005]{}
Stern, D., Eisenhardt, P., Gorijan, V., et al. 2005, ApJ, 
631, 163

\bibitem[Vallenari et al. 2005]{}
Vallenari, A., Schmidtobreick, L., \& Bomans, D. J. 2005, 
A\&A, 435, 821

\bibitem[van den Hoek et al. 2000]{}
van den Hoek, L. B., de Blok W. J. G., van der Hulst, J. M., 
\& de Jong, T. 2000, A\&A, 357, 2000

\bibitem[Vazquez \& Leitherer 2005]{}
Vazquez, G. A., \& Leitherer, C. 2005, ApJ, 621, 695 

\bibitem[Walterbos \& Schwering 1987]{}
Walterbos, R. A. M., \& Schwering, P. B. W., 1987, A\&A, 180, 27


\bibitem[Warner et al. 2004]{}
Werner, M., Roellig, T. L., Low, F. J., et al. 2004, ApJS, 
154, 1

\bibitem[Wu et al. 2006]{}
Wu, Y., Charmandaris, V., Hao, L., Brandl, B. R., Bernard-Salas, J., 
Spoon, H. W. W., \& Houck, J. R. 2006, ApJ, 639, 157

\bibitem[Wynn-Williams \& Becklin 1974]{}
Wynn-Williams, C. G. \& Becklin, E. 1974, PASP, 86, 5 

\bibitem[Zackrisson et al. 2005]{}
Zackrisson, E., Bergvall, N., \& {O}stlin, G. 2005, A\&A, 435, 
29
\end{thebibliography}
\end{document}